\documentclass[12pt]{article}
\usepackage{graphicx}
\usepackage{natbib}
\usepackage{amssymb,amsmath,color,xcolor}
\usepackage{bm}
\usepackage{booktabs}
\usepackage[affil-it]{authblk}
\usepackage{url}
\usepackage[toc,page]{appendix}
\usepackage[color=lightgray]{todonotes}
\usepackage[bf,font={small,sl}]{caption}

\renewcommand{\baselinestretch}{1.2}
\usepackage[margin=1in]{geometry}

\title{\textbf{Estimation and simulation of foraging trips in land-based marine predators}} 
\author{Th\'eo Michelot$^1$\footnote{tmichelot1@sheffield.ac.uk (All supplementary material is available on request.)}, Roland Langrock$^2$, Sophie Bestley$^{3,4}$, Ian D.\ Jonsen$^5$, Theoni Photopoulou$^{6,7}$, Toby A.\ Patterson$^8$}
\affil{$^1$University of Sheffield, $^2$Bielefeld University, $^3$Australian Antarctic Division, $^4$Institute for Marine and Antarctic Studies, $^5$Macquarie University, $^6$Nelson Mandela Metropolitan University, $^7$University of Cape Town, $^8$CSIRO Oceans and Atmosphere}
\date{}

\begin{document}
\maketitle

\vspace{-5mm}
\begin{abstract}
The behaviour of colony-based marine predators is the focus of much research globally. Large telemetry and tracking data sets 
have been collected for this group of animals, and are accompanied by many empirical studies that seek to segment tracks in some useful way, as well as theoretical studies of optimal foraging strategies. 
However, relatively few studies have detailed statistical methods for inferring behaviours in central place foraging trips. In this paper we describe an approach based on hidden Markov models, which splits foraging trips into segments labeled as ``outbound'', ``search'', ``forage'', and ``inbound''. By structuring the hidden Markov model transition matrix appropriately, the model naturally handles the sequence of behaviours within a foraging trip. Additionally, by structuring the model in this way, we are able to develop realistic simulations from the fitted model. We demonstrate our approach on data from southern elephant seals (\emph{Mirounga leonina}) tagged on Kerguelen Island in the Southern Ocean. We discuss the differences between our 4-state model and the widely used 2-state model, and the advantages and disadvantages of employing a more complex model.  
\end{abstract}

\section{Introduction}
\label{S:Intro}

Central place foraging (CPF) is a widely applied concept in ecology \citep{olsson2014model, higginson2015influence}. Many terrestrial species with home ranges, or shelters, can be regarded as central place foragers \citep{bell1990central}. Colony-based marine animals, such as seabirds and seals, that must moult, breed and raise young on land or ice can also use CPF strategies. It has been hypothesized that, due to density dependence, waters near to the colony may become depleted of prey \citep{ashmole1963regulation}, or simply that the most profitable prey are spatially separated from land-based colonies, necessitating trips to more distant foraging areas \citep{Oppel2015}. In many seabirds, during phases when the young are being fed and reared, adult birds are constrained to forage within closer range of the colony \citep{boyd2014movement,patrick2014individual}. Pinnipeds must also haul out to periodically moult, as well as for mating and rearing young \citep{russell2013uncovering}. These animals are the subject of considerable ongoing research, often utilising tracking techniques to collect movement data at sea \citep{hays2016key}. Moreover, many are the subject of intense conservation efforts \citep{lonergan2007using, croxall2012seabird, hamer2013endangered, martin2015reducing, jabour2016marine}. 

Often, studies aim to assess how CPF animals apportion their time between searching and active foraging, and to identify particular characteristics, for example relating to habitat usage, trip length, activity budgets, and other variables \cite[e.g.][]{staniland2007energy, raymond2015important,hindell2016circumpolar,Patterson2016}. How these quantities vary with ontogenetic stage or age is often important; for example, naive young animals versus experienced adults, or sex-specific foraging strategies \citep[e.g.][]{breed2009sex,hindell2016circumpolar}. To evaluate hypotheses about movements in CPF, it is very helpful to have models which objectively classify movement into different modes (or ``phases''), with different statistical properties, indicating differences in underlying behaviour \citep{langrock2012flexible}.

Several recent CPF studies have looked at switching models which include latent behavioural states, but usually these are general models that are not specifically designed to capture the behavioural cycle of CPF \citep[e.g.,][]{breed2009sex, jonsen2013state, jonsen2016joint}. Latent states are often given labels such as ``transit'' and ``resident''. In the latter, the notion of area-restricted search \citep{Kareiva:Odell:1987, fauchald2003using, morales2004extracting} is often invoked. In an area with high foraging returns, the animal is expected to undertake less directed movements, and therefore higher turning rates, and generally lower speeds of travel. These concepts have proven useful for modelling movements of CPF, but explicitly incorporating the trip structure in the model would give a more nuanced understanding of the likely behavioural sequences. At the most basic level, an animal must leave a colony, transit out to (one or more) foraging grounds, search and obtain food, and then eventually return.

Our aim here is to construct a model which captures the following sequence of movement modes: outbound $\rightarrow$ search $\rightarrow$ forage $\rightarrow \ldots \rightarrow$ search $\rightarrow$ inbound. We outline the random walk models which can be used to represent these movement modes, and then show how these can be integrated into the hidden Markov model (HMM) framework. HMMs have been used widely in animal movement modelling \citep{langrock2014modelling, mckellar2015using, deruiter2017multivariate, leos2016analysis, auger2016evaluating}. They represent a computationally efficient approach for fitting models with discrete latent states to time series data. A thorough description of HMMs for animal movement is given in \citet{langrock2012flexible}. 

An important subsequent focus of the model we seek to construct is that it should be suitable for simulation of foraging trips. Simulation is already used in habitat modelling of central place foragers as null models for distribution that account for some areas being inaccessible due to distance constraints \citep{raymond2015important,wakefield2009wind}. However, estimation of such simulation models is often ad-hoc \citep{matthiopoulos2003model}. Beyond the specific case of habitat modelling, even simulation models which capture only a few aspects of foraging behaviour would be useful in making predictions, with associated uncertainty, from finite samples of individuals drawn from populations. The desire to simulate from the fitted model places a requirement of greater realism on the movement model structure. This further motivates the development of the trip-based movement model for CPF.

We acknowledge from the outset that the model we present below is a simplification of the true processes under investigation. Despite the broader applicability of the CPF concept, we restrict our usage to the situation of marine predators that are constrained to return to land after a certain period of time. It should be noted however that the methods we present may well have broader application. CPF is likely to be influenced by patchiness operating at scales which are not observable from the telemetry data, and dependent on local conditions (only some of which may be observable via remote sensing). To fully generalize our methods, inclusion of environmental data will be necessary \citep{labrousse2015winter}. Considerations of these ideas have tended to employ computationally demanding techniques \citep[e.g.][]{mcclintock2012general}, which can limit their applicability with large data sets. So far, models for large data sets (i.e.\ many observed locations) have been underrepresented.

Below, we describe the modeling framework and demonstrate estimation using data from southern elephant seals (\emph{Mirounga leonina}). We then show simulation of foraging trips from the fitted model, and assess which aspects of the real trips are replicated well and which are not. Finally, we discuss how a model of this type can be tailored to a given species' case and extended to include other covariates (e.g.\ environmental) that may influence movement behaviour. Additionally, we discuss the limitations of our model, and how these relate to the general problem of simulation and subsequent prediction of behaviour from fitted multistate movement models.   

\section{Methods}
\label{S:Methods}

\subsection{Building blocks for the overall model}
\label{S:RW}
\begin{figure}[!htb]
\centering
\includegraphics[width=0.6\textwidth]{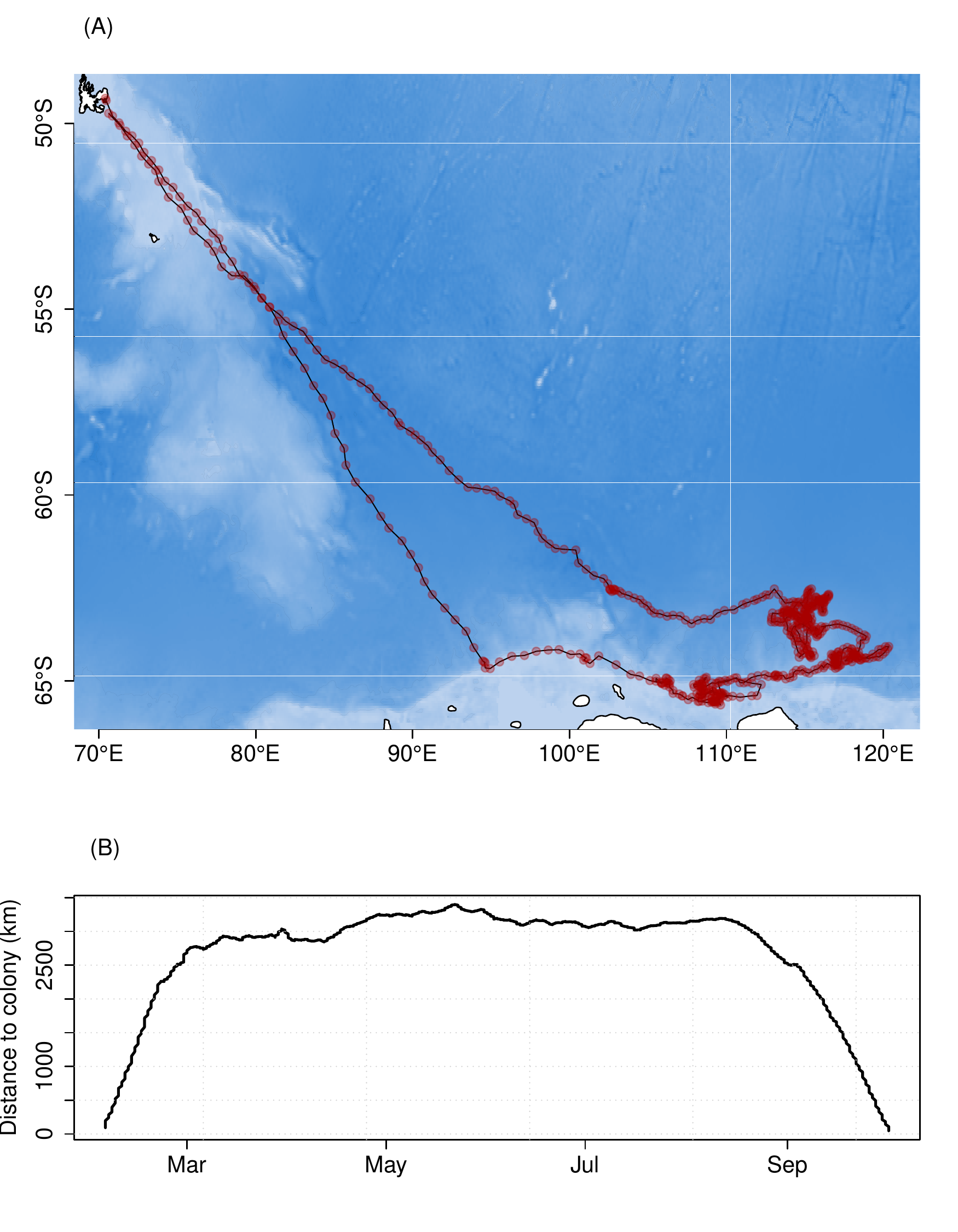}
\caption{(A) An example track of a southern elephant seal tagged at Kerguelen Island (top left), which makes an outward foraging trip to the Antarctic continental shelf before returning. (B) The distance from Kerguelen Island through time, clearly showing rapid outward transit, foraging movements over an extended period of time, and the rapid return transit.}
\label{example_track}
\end{figure}

We first describe the types of random walks required to describe the different segments of a foraging trip, and then how these are combined within the full model, an HMM. Figure \ref{example_track} illustrates the typical pattern of an elephant seal's foraging trip. The maps were produced with the R package marmap \citep{pante2013marmap}. 

The trip consists of (at least) three clearly distinct phases of movement: the fast and directed trip from the colony to the sea ice region, a period of slower and less directed movement near the ice, and the fast and directed trip from the ice to the colony. Marine prey resources are patchily (non-uniformly) distributed at multiple spatial scales \citep{fauchald2006hierarchical}, so animals will typically still need to search between dynamic favourable forage patches within and around the sea ice and Antarctic shelf regions \citep[see for example inset panels of Figure 2 in][]{bestley2013integrative}. This search may manifest as slower, less directed movement than the migration transits, but faster and more directed movement than area restricted search

We choose the following four movement modes for our analysis, for the distinct phases within a trip: (1) outbound trip, away from the central place (here, the colony at Kerguelen Island) and towards the foraging regions, characterized by very fast and highly directed movement; (2) search-type movement in the foraging regions, characterized by moderately fast movement and some directional persistence (though without a clear destination); (3) foraging activities, characterized by non-directed, slow movement; and (4) return trip, back to the central place, with very fast and again highly directed movement. We label the movement modes for convenience, but note that each phase of the trip (in particular ``search'' and ``forage'') can potentially encompass several behaviours, perhaps even behaviours that are functionally equivalent, like different types of foraging behaviour, as long as they lead to similar movement patterns.

Both (2) and (3) can be adequately modelled using commonly applied correlated random walks (CRWs). CRWs involve correlation in directionality, and can be represented by modelling the turning angles of an animal's track using a circular distribution with mass centred either on zero (for positive correlation) or on $\pi$ (for negative correlation). For example, we could model phase (2), which involves positive correlation in directionality, by assuming
\begin{equation*}
\text{bearing}_t \sim \text{von Mises}\,(\text{mean}=\text{bearing}_{t-1},\ \text{concentration}=\kappa),
\end{equation*} 
with $\kappa$ denoting a parameter to be estimated. (Larger $\kappa$ values lead to lower variance of the von Mises distribution, and hence higher persistence in direction.) Alternatively, the mean can be specified to be $\text{bearing}_{t-1}-\pi$, corresponding to an expected reversal in direction, which is often found in encamped, foraging, or resting modes (e.g.\ ``encamped'' behaviour in \citealp{morales2004extracting}, and ``area-restricted search'' in \citealp{towner2016sex}). Instead of specifying the mean {\em a priori}, it can also be estimated from the data. That is, we can consider
\begin{equation*}
\text{bearing}_t \sim \text{von Mises}\,(\text{mean}=\text{bearing}_{t-1} + \lambda,\ \text{concentration}=\kappa),
\end{equation*} 
where $\lambda \in [-\pi,\pi)$ is a parameter to estimate. We note that CRWs can alternatively be constructed by modelling turning angles rather than bearings, leading to equivalent model formulations, with
\begin{equation*}
\text{angle}_t \sim \text{von Mises}\,(\text{mean}= \lambda,\ \text{concentration}=\kappa).
\end{equation*} 
Choosing between these two formulations is only a matter of convenience.

For modelling phases (1) and (4), biased random walks (BRWs) are better suited. Bias in random walks usually (though not necessarily) refers to a tendency towards (or away from) a particular location, sometimes called a centre of attraction (or point of repulsion). For example, a bias towards a location $(c_1, c_2)$ is obtained by assuming that 
\begin{equation*}
\text{bearing}_t \sim \text{von Mises}\,(\text{mean}=\phi_t,\text{concentration}=\kappa),
\end{equation*}
where $\phi_t = \arctan \left[ (y_c - y_t)/(x_c - x_t) \right]$ is the direction of the vector pointing from the animal's current position, $(x_t, y_t)$, to the central place, $(x_c, y_c)$. 

In principle, it is also possible to construct random walks that are both biased and correlated (BCRWs), thereby trading off directional persistence and possible turning towards the destination. However, in our experience, it is difficult to statistically distinguish CRWs and BRWs. As a consequence, usage of BCRWs tends to lead to numerical instability within the model fitting procedure, as it is challenging to estimate the respective weights of the biased and correlated components of the process. For more details on CRWs, BRWs and BCRWs, see \citet{Codling2008}.

As can be seen in Figure \ref{example_track}, the animal tends to remain within each phase of movement for some time before switching to a different phase. To accommodate both the persistence and the stochastic switching between different phases of the movement, we use an underlying (unobserved) state process. The state process, $S_1\ldots,S_T$, takes values in $\lbrace 1,\dots,N \rbrace$, such that, at each time $t$, the movement observed follows one of $N$ types of random walks (possibly correlated and/or biased), as determined by the current state. We take the state process to be a Markov chain, which together with the observation process, i.e.\ the BRWs and CRWs conditional on the current state, defines a hidden Markov model for the animal's movement (HMM; see Chapter 18 in \citealp{zucchini2016hidden}). The state process is characterized by the transition probabilities
\begin{equation*}
  \gamma_{ij} = \Pr(S_t=j \vert S_{t-1}=i),\ \text{ for } i,j \in \lbrace 1,\dots,N \rbrace .
\end{equation*}

In the trip-based movement detailed above, there are $N=4$ states, corresponding to the four phases of movement. Note that, in general, there is no guarantee of a one-to-one equivalence between the (data-driven) states of the Markov chain and the actual behaviours of the animal \citep{Patterson2016}. As such, the states should be interpreted with care, but they are often useful proxies for the behavioural modes. In subsequent sections, we use the terms ``behaviours'' and ``states'' interchangeably.

Framing the model as an HMM makes it possible to use the very efficient HMM machinery to conduct statistical inference. In particular, the model parameters can relatively easily be estimated by numerically maximizing the likelihood, which can be calculated using a recursive scheme called the forward algorithm \citep{zucchini2016hidden}. The main challenge herein lies in identifying the global rather than a local maximum of the likelihood --- the same or similar problems arise when using the expectation-maximization algorithm or Markov chain Monte Carlo sampling to estimate parameters. The efficiency of the forward algorithm is one of the key reasons for the popularity and widespread use of HMMs, and similar algorithms can be applied for forecasting, state decoding, and model checking. In particular, we use the Viterbi algorithm to decode the most likely sequence of underlying states in Section \ref{S:viterbi}. More details on inference in HMMs are given in \cite{zucchini2016hidden} and, for the particular case of animal movement modelling, in \cite{Patterson2016}. For our case study, we used the R optimization function \texttt{nlm} to numerically maximize the likelihood, which was partly written in C++ for computational speed.

\subsection{Case study data}
The southern elephant seal (\emph{Mirounga leonina}) is a pinniped top predator with a circumpolar distribution throughout high latitudes in the southern hemisphere \citep{carrick1962studies}. Haul out phases on land occur at fairly predictable times during the annual cycle for moulting and breeding \citep{hindell1988seasonal}. The species is known for making large scale migrations from isolated subantarctic land colonies, both southwards to the sea ice zone around the Antarctic continental margin, and also into open ocean pelagic zones \citep{biuw2007variations,labrousse2015winter,hindell2016circumpolar}. A recent study \citep{hindell2016circumpolar} applied two-state movement models (\emph{sensu} \citealp{morales2004extracting}) to a large data set of several hundred individual tracks. This study was focused on detecting basin-scale patterns in foraging effort, rather than explicitly modelling the sequence of behaviours within individual foraging trips. 

In this case study, we examine trips from 15 animals (8 adult females and 7 subadult males) tagged at Kerguelen Island. These animals were fitted with telemetry units of the Sea Mammal Research Unit (St Andrews, UK) which transmit data via the Argos satellite network \citep{photopoulou2015generalized}. Frequently, the elephant seals remained at the colony for extended periods at the start of the time series, and similarly after returning from foraging trips. Because of the HMM structure, the highly stationary data from these periods holds little information and rather serves to leave potential for numerical instability. Data from these periods were removed prior to use in the HMM. Near stationary periods immediately prior to colony departure and/or following return to the colony were identified for removal by examining one-dimensional time series plots of the Argos longitude and latitude observations. The truncated trips lasted between five and eight months, except for two incomplete trips, during which the data collection was interrupted before the animals returned to the colony (tags may fail before trip completion).

Due to the occasional large errors and the irregular timing of the Argos location observations, these data were filtered using a state-space model \citep[SSM,][]{jonsen2013state} to obtain a regular time sequence of location estimates with reduced uncertainty. The SSM used was a variant of that described in \citet{Jonsen2005}, implemented with the R package TMB \citep{Kristensen2016}. Associated R and C++ code, for pre-processing the Argos data, are available on Github, at \url{github.com/ianjonsen/ssmTMB}. The state-space model was fit with a 2.4-hour time step, yielding ten location estimates per day. Other time steps were evaluated but 2.4 h provided the best fit according to AIC and auto-correlation functions of the residuals. This time step resulted in calculated step lengths (speeds) and turning angles (or, equivalently, bearings) that had relatively low contrast between movement phases apparent in the observed data.  Accordingly, we sub-sampled the estimated locations to every fourth time step (i.e. 9.6 hour frequency). The data provided to the HMM were step lengths and turning angles, as described in Section \ref{S:RW}.

\subsection{Model details}
\label{S:Model}
For the elephant seal case study, we employ the structure of a four-state HMM as described in Section \ref{S:RW}. The first state corresponds to the outbound trip from the colony to a foraging region, and is modelled by a BRW with repulsion from the colony. The animal then alternates between states 2 (``search'') and 3 (``forage''), each using a CRW. Finally, the process switches to the inbound trip, modelled by a BRW with attraction towards the colony. The movement is measured in terms of step lengths and turning angles -- modelling the latter is equivalent to modelling bearings, but here easier to implement. We use a gamma distribution to model the step lengths in each state, and a von Mises distribution for the turning angles. The mean turning angles are estimated for both state-dependent CRWs (in states 2 and 3, respectively), instead of being fixed \emph{a priori}. This results in fourteen parameters to estimate at the level of the observation process: four shape and four scale parameters (for the gamma-distributed steps), plus four concentration parameters and two means (for the angles). For states 1 and 4, no mean parameter needs to be estimated for the associated BRW, as the expected direction is determined through the bias. 

Using the notation introduced in Section \ref{S:RW}, we write the transition probability matrix as
\begin{equation*}
  \bm{\Gamma} = \begin{pmatrix}
    \gamma_{11} & \gamma_{12} & 0 & 0 \\
    0 & \gamma_{22} & \gamma_{23} & \gamma_{24} \\
    0 & \gamma_{32} & \gamma_{33} & 0 \\
    0 & 0 & 0 & 1
  \end{pmatrix}.
\end{equation*}
      
This structure ensures that the sequence of states follows the behavioural cycle described in the Introduction, by preventing some transitions. Note that we could choose $\gamma_{13}$ and $\gamma_{34}$ to be non-zero, to allow the process to switch from outbound to forage, and forage to inbound, respectively. In this case study, we decided to prevent these transitions, i.e.\ we assumed a transitional regime of moderately fast and directed movement (search) between a phase of fast and directed movement (outbound or inbound), and a phase of slow non-directed movement (forage). 

This formulation also leads to improved numerical stability of the estimation, as it reduced the number of parameters to estimate. Here, state 4 is an absorbing state, as we only consider tracks comprising (at most) one trip away from the colony. It would be straightforward to relax this constraint, e.g.\ by choosing $\gamma_{41}>0$. For tracks comprising several years of data, a fifth state could be added to model the movement of the seals at the colony.

In analyses like this one, it is often of interest to understand the drivers of behavioural switches, by expressing the transition probabilities as functions of time-varying covariates \citep[see, e.g.,][]{mckellar2015using, Breed2016}. Here, we introduce two covariates: the great-circle distance to the colony from the location at time $t$, $d_t$, and the time since departure from the colony, $t-t_0$. These are included for two key aspects which are both relevant for CPF foraging behaviour and are also necessary to construct a model which will replicate trips in a simulation setting. Specifically, we need to model the fact that animals often make fast directed trips away from the colony and tend to switch into other movement modes once they have reached foraging grounds. In this case we use distance from the colony as a covariate affecting the probability of transitioning from outbound to search,

\begin{equation*}
  \gamma_{12}^{(t)} = \text{logit}^{-1} \left( \beta_0^{(12)} + \beta_1^{(12)} d_t \right).
\end{equation*}

Somewhat similarly, animals cannot remain at sea indefinitely. Therefore, we use time since departure  from the colony as a covariate on the switch from search to inbound ($\gamma_{24}$). As there are two non-zero probabilities of switching out of state 2, we use a special case of the multinomial logit link, the general expression of which is given e.g.\ in \cite{Patterson2016}. In our model,
\begin{align*}
  \gamma_{23}^{(t)} = \dfrac{\exp\bigl(\beta_0^{(23)}\bigr)}{1+\exp\bigl(\beta_0^{(23)}\bigr) + \exp(\eta_{24})}, \text{ and }
  \gamma_{24}^{(t)} = \dfrac{\exp(\eta_{24})}{1+\exp\bigl(\beta_0^{(23)}\bigr) + \exp(\eta_{24})},
\end{align*}
where
\begin{equation*}
  \eta_{24} = \beta_0^{(24)} + \beta_1^{(24)} (t - t_0).
\end{equation*}

The $\beta_k^{(ij)} \in \mathbb{R}$ are parameters to be estimated. Note that, because the rows of the transition probability matrix must sum to 1, $\gamma_{11}$ and $\gamma_{22}$ are also time-varying in this example. There are six parameters to estimate in the state-switching process: the five $\beta_k^{(ij)}$ coefficients and $\gamma_{32}$. The remaining elements of the matrix, i.e.\ $\gamma_{11}$, $\gamma_{22}$ and $\gamma_{33}$, are obtained from the row constraints.

This results in a total of twenty parameters to estimate: fourteen parameters for the state-dependent distributions of steps and angles, and six parameters for the transition probabilities.

\subsection{Simulation from fitted model}
\label{S:sim}
Having estimated model parameters from the real data, it is possible to simulate movement from the model described in Section \ref{S:Model}. A simulated track starts near Kerguelen Island, in state 1 (outbound trip). The bearing is initialized from a von Mises distribution with a mean pointing towards the South, to mimic the elephant seals' movement. The directionality of the movement in state 1 ensures that the trajectory goes southward, overall. At each time step, the state process is simulated from the estimated (possibly time-varying) switching probabilities. Then, a step length and a bearing are simulated from the estimated gamma and von Mises distributions, respectively. The corresponding longitude and latitude coordinates are derived using the R package geosphere \citep{hijmans2016}. The new location is rejected if it is on land, using the borders defined in the data set \texttt{wrld\_simpl} of the  maptools R package \citep{bivand2016}. The track ends once the trajectory is back at the colony. In practice, we chose to stop the simulation once a location is simulated within a 20-kilometre radius around Kerguelen Island.

\section{Results}
The model described in Section \ref{S:Model} was fitted to 15 elephant seal tracks, each corresponding to one individual trip away from the colony. The tracks comprise about 7300 locations, and it took around one minute to fit the model on a dual-core i5 CPU.

We include the code used to fit the model in the supplementary material. Note that, for speed, we implemented the likelihood function in C++, using Rcpp \citep{eddelbuettel2011rcpp}. In the supplementary material, we also provide the data set comprising the 15 tracks.

\subsection{Estimated turn and step-length distributions}
\begin{figure}[htb]
\centering
\includegraphics[width=\textwidth]{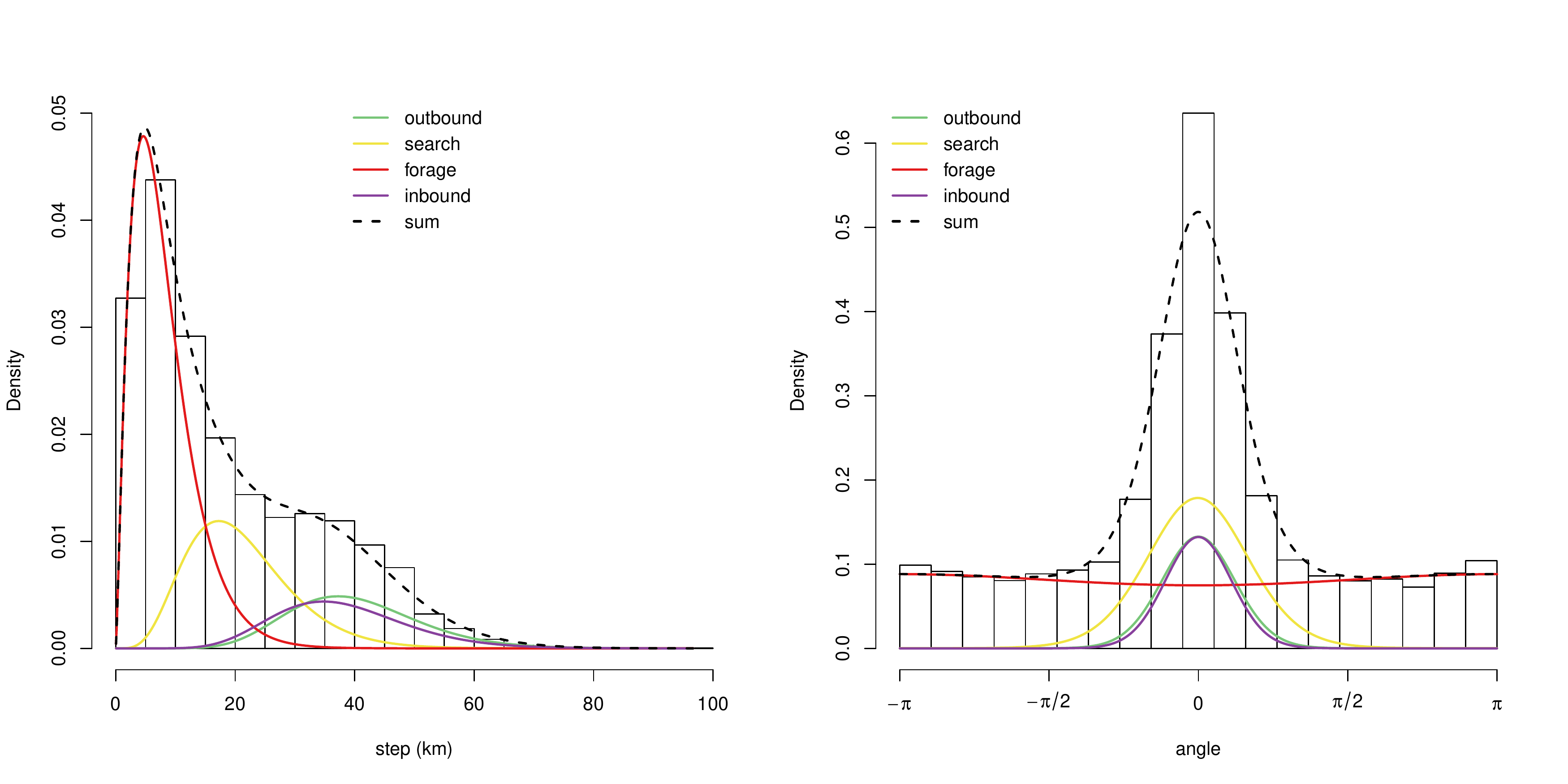}
\caption{Estimated state-dependent distributions for the step lengths (left) and the turning angles (right).}
\label{F:densities}
\end{figure}

Figure \ref{F:densities} shows histograms of the step lengths and turning angles of the data, on which are plotted the estimated state-dependent gamma and von Mises densities. The state-dependent densities for each state here have been weighted according to the proportion of time the corresponding state is active, as determined using the Viterbi algorithm \citep{zucchini2016hidden}. Similarly, for both the step lengths and the turning angles, Figure \ref{F:densities} also displays the cumulative distribution, i.e.\ the sum of these weighted densities. Based on visual inspection, these cumulative distributions do not indicate any lack of fit of the model. 

Formal model checks can be conducted using forecast pseudo-residuals, which use the probability integral transform to effectively compare the observation at each time $t$ to the associated forecast distribution based on the observations up to time $t-1$. In case of the turning angles, the definition of the pseudo-residuals is essentially arbitrary due to the circular nature of this variable (cf.\ \citealp{langrock2012flexible}). Thus, we restrict the model check to the step length variable. The quantile-quantile plot of the pseudo-residuals for the step lengths, against the standard normal distribution, is shown in Appendix A4.
A few of the tracks include steps of slow movement near the colony. They are not captured by the ``outbound'' and ``inbound'' states of fast movement, such that they appear as outliers in the qq-plot; this could be resolved during the preprocessing, by excluding the corresponding observations. The model also slightly underestimates the number of long steps (roughly between 40 and 50 km). In this regard, the fit could be improved by using more flexible step length distributions, albeit at a computational cost \citep{langrock2015nonparametric}. However, note that the improvement in inference on the state-switching dynamics would be minimal.

The states corresponding to the outbound and inbound movements display very similar features, with high step lengths and strong directional persistence; the distinction is that the colony acts as a centre of repulsion in the former, and a centre of attraction in the latter. The foraging phases are characterized by shorter steps, i.e.\ slower movement, and less directional persistence, with a roughly flat distribution of turning angles. In the search-type movement mode, the model captures moderately long steps and directed movement, making it clearly distinct from foraging behaviour.

The estimates of all the model parameters are provided in Appendix A2, in the supplementary material.

\subsection{Estimated state sequences }
\label{S:viterbi}
The most probable state sequence was computed with the Viterbi algorithm. Figure \ref{F:alltracks} shows the 15 tracks, coloured by decoded states. The individual decoded tracks are provided in Appendix A1. In all tracks, the first state corresponds to the animal moving quickly towards the south. Then, the behaviour alternates between searching (state 2) and foraging (state 3) periods, typically near the ice or in Antarctic continental shelf waters. In general, more northerly search behaviour is apparent in the westernmost tracks. Eventually, the animal switches to state 4 as it starts moving back towards the island colony. 

\begin{figure}[htb]
\centering
\includegraphics[width=\textwidth]{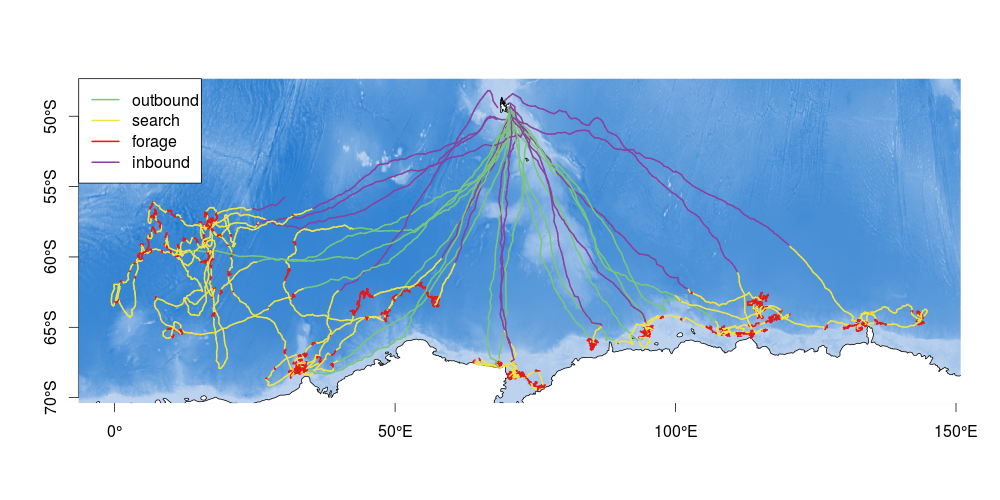}
\caption{Fifteen elephant seal tracks, coloured by Viterbi-decoded states. The white area at the bottom is the Antarctic continent.}
\label{F:alltracks}
\end{figure}

Overall, the model appears to adequately identify the outbound and inbound trips. However, we suspect that the decoded state sequence might sometimes fail to capture the exact timing of the transitions out of the outbound state, and the transitions into the inbound state. In some tracks, the animal goes through a transitional phase, between the outbound trip and search behaviour, in which the movement is slower but still very directed. Although these periods are still arguably part of the outbound trip, they might be attributed to the searching state, due to the decrease in speed. The same situation arises during the transition from search to inbound trip.

\begin{figure}[htb]
\centering
\includegraphics[width=\textwidth]{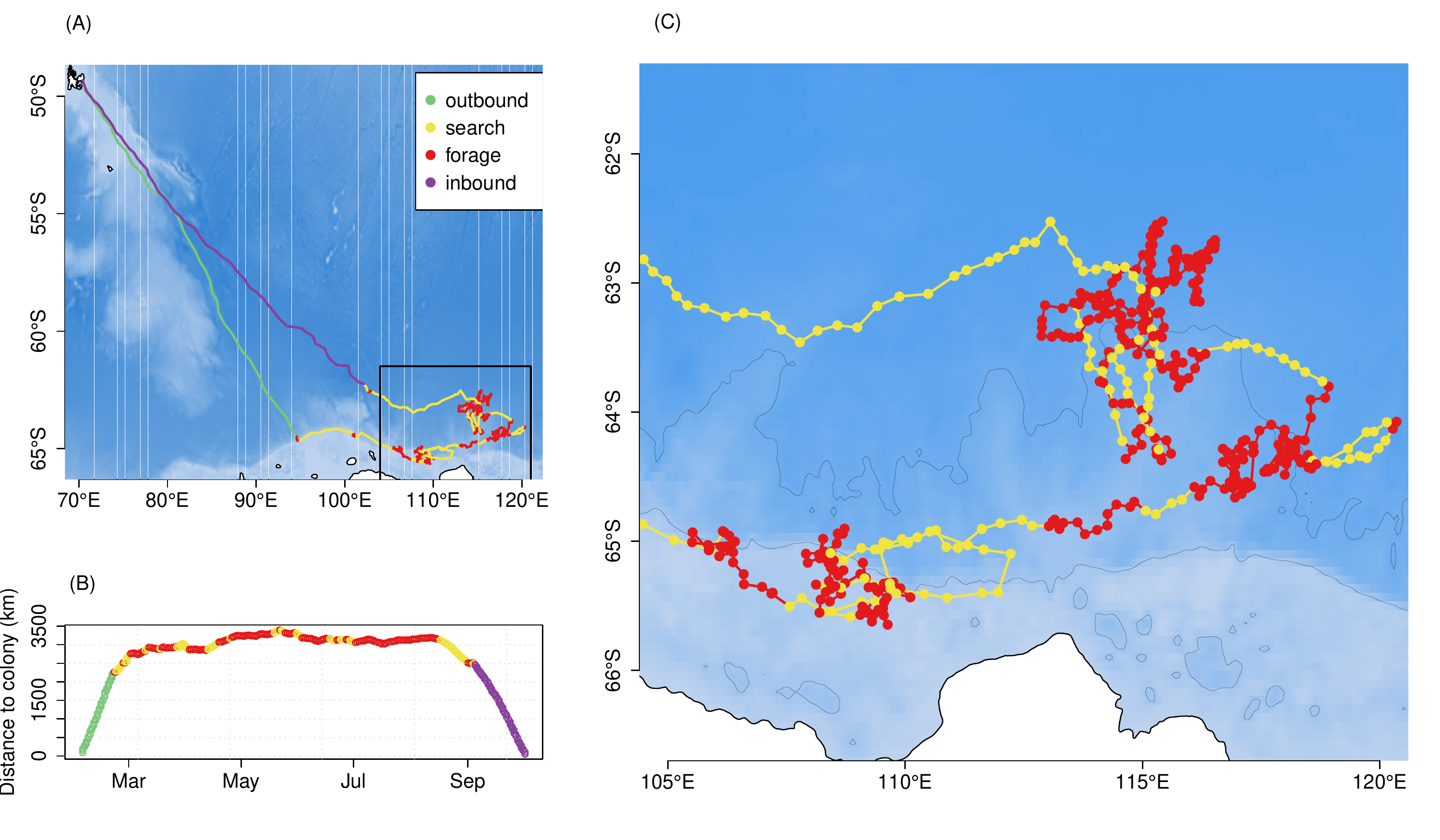}
\caption{(A) An example track of a southern elephant seal. (B) The distance from Kerguelen Island through time. (C) A `zoomed in' part of the track shown in (A). The four colours correspond to the most probable states, decoded with the Viterbi algorithm. (A) and (B) clearly illustrate the phases of fast and directed movement, which are attributed to the outbound and inbound trips. (C) shows in more detail the different patterns in the animal's movement, when near the sea ice region, which distinguish between search and foraging behaviours.}
\label{F:decodedtrack}
\end{figure}

Figure \ref{F:decodedtrack} demonstrates state decoding more specifically, on the track presented in Figure \ref{example_track}. In particular, sub-Figure \ref{F:decodedtrack}(C) shows how the localized movements of the elephant seal near the Antarctic continent is split into two very distinct behaviours, which seem to be apportioned adequately between states 2 and 3.

\subsection{Estimated effects of covariates}

\begin{figure}[htb]
\centering
\includegraphics[width=\textwidth]{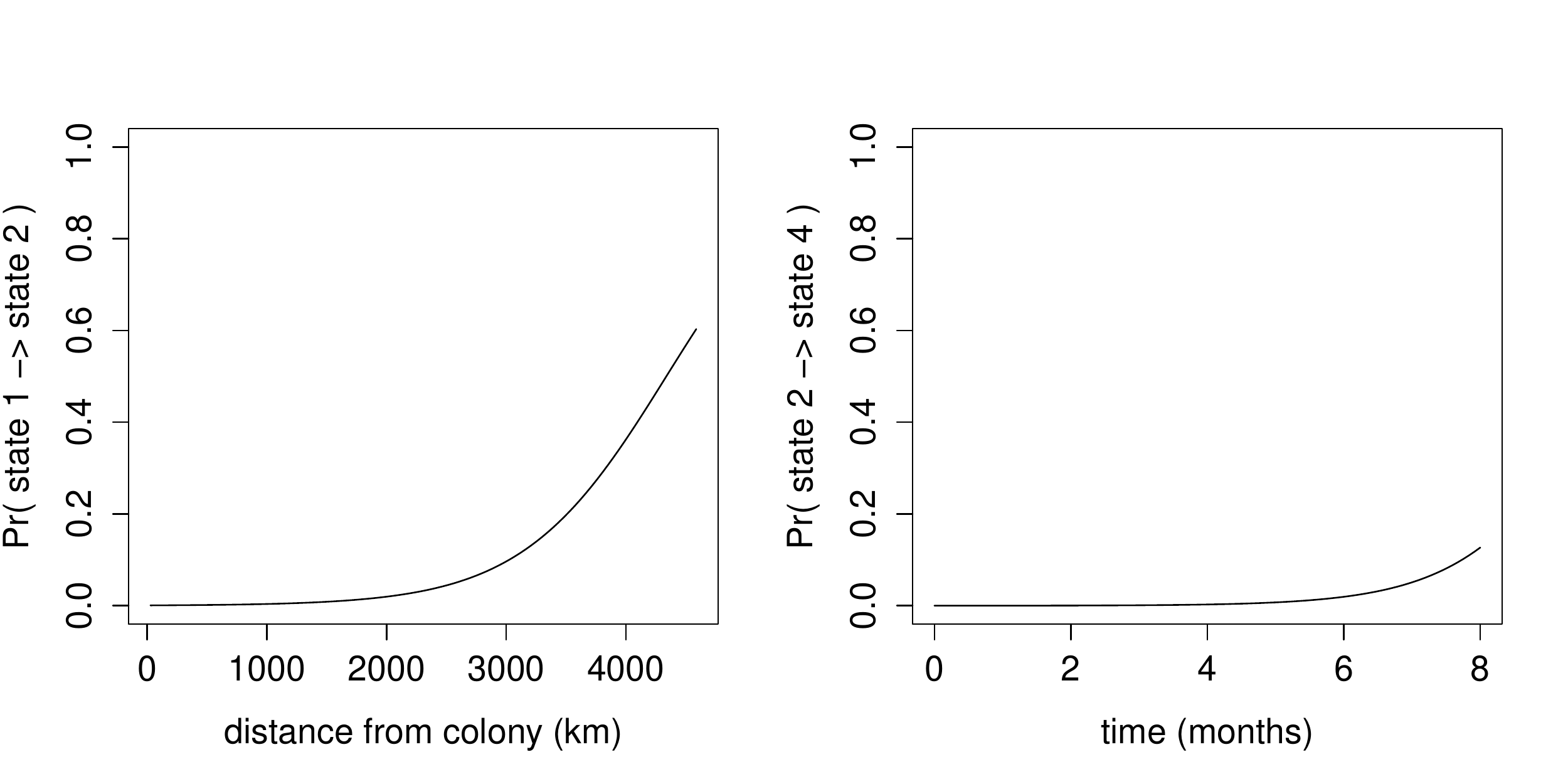}
\caption{Transition probabilities from outbound to search as a function of the distance from the colony (left), and from search to inbound as a function of the time spent away from the colony (right).}
\label{F:tpm}
\end{figure}

The transition probabilities were estimated as functions of the distance to the colony, and the time spent away from the colony, as described in Section \ref{S:Model}. Figure \ref{F:tpm} displays plots of the transition probabilities from state 1 to state 2 (end of outbound trip), and from state 2 to state 4 (start of inbound trip).

The HMM predicted that elephant seals were unlikely to switch away from state 1 when close to the colony, but the probability increased quickly at distances greater than 3000 km, when the animals tend to start searching for foraging patches. Moreover, during the first few months away from the colony, elephants seals do not switch to state 4 (return trip), but instead tend to cycle through search and foraging phases. Later, after about six months, the probability of switching to state 4 starts to increase. 

This is consistent with the annual cycle in this species and the timing of return to the island colony after the long post-moult migration \citep{slip1999population, hindell1988seasonal,mccann1980population}. Therefore, the estimated relationships between covariates and behaviour is consistent with the known behaviour of  elephant seals and their annual moulting and breeding cycles. 

\subsection{Simulation results}
\begin{figure}[htb]
\centering
\includegraphics[width=\textwidth]{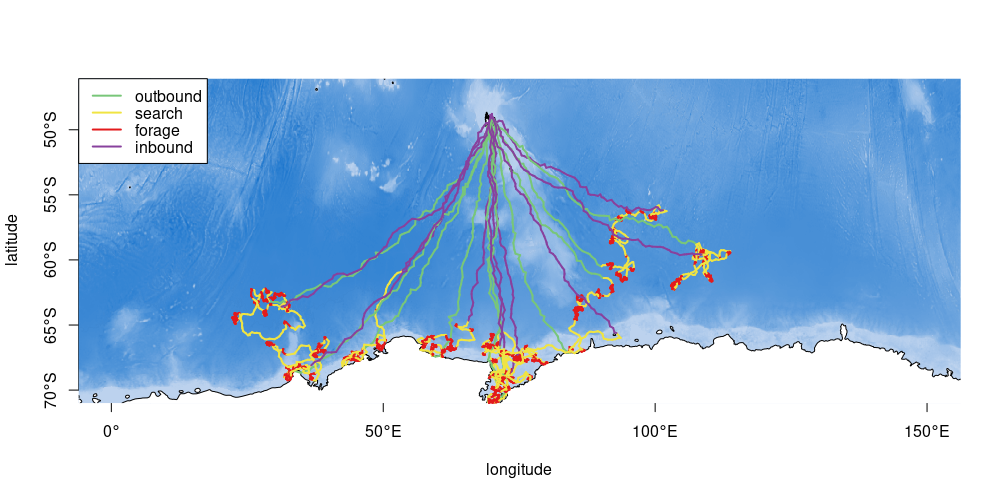}
\caption{Ten simulated movement tracks, obtained with the MLE of the parameters of the fitted model.}
\label{F:simtracks}
\end{figure}

One of the main advantages of our approach over simpler HMMs (e.g.\ HMMs with fewer states, no constraints on the switching probabilities, only based on CRWs...) is the possibility to simulate realistic movement tracks from the fitted model. The simulation procedure is described in Section \ref{S:sim}. Figure \ref{F:simtracks} shows ten tracks, simulated from the fitted model.

The simulated tracks display many of the features of the real ones. To compare them, we simulated 100 tracks, and summarized the proportion of time allocated to each state, and the mean duration spent in each behaviour (before switching to another behaviour), in Table \ref{T:stateprop}. In the simulated data, we find the overall proportional state allocation very well represented. However, we find on average slightly shorter behavioural phases for states 2 and 3 (search and forage) than in the real data. This may partly be due to the assumed Markov property of the state process which, for states not affected by covariates (e.g.\ the foraging state), implies that the times spent within the state are geometrically distributed \citep{zucchini2016hidden}. We suggest ways to relax this assumption in Section \ref{S:prediction}. Histograms of the dwell times in each state, for the real and simulated tracks, are provided in Appendix A3.

\begin{table}[htb]
\centering
\caption{Comparison of the real tracks and 100 simulated tracks, in terms of overall proportion of observations attributed to each state, and of mean dwell time in each behaviour (in days).}
\begin{tabular}{lcccccc}
	& & \multicolumn{2}{c}{Overall proportion} & & \multicolumn{2}{c}{Mean state dwell time} \\
	& & Real data & Simulated data & & Real data & Simulated data \\
    \midrule
    State 1 & & 12.9\% & 11.2\% & & 25.4 & 23.2 \\ 
    State 2 & & 24.2\% & 23.3\% & & 4.5 & 3.6  \\
    State 3 & & 51.1\% & 53.7\% & & 10.4 & 9.0 \\
    State 4 & & 11.7\% & 11.8\% & & 24.6 & 24.4
\end{tabular}
\label{T:stateprop}
\end{table}

The simulation successfully captures the southward direction of the outbound trip. Then, the process switches to search and forage behaviours at a realistic distance to the colony, as the probability of this transition is a function of the distance to Kerguelen Island. The extent of movements within each state, which is informed by the estimated step and angle distributions, is also realistic. However, the spatial distribution of foraging activity is not tied to correspond to what is observed in real tracks. Thus, the exact locations of searching and foraging activities in the simulated data are of no environmental relevance. This could be improved by including environmental covariates; this is discussed in more detail in Section \ref{S:prediction}.

The durations of simulated trips are reasonable for the study species: out of the 100 simulated trips, 90 lasted between five and nine months.

\section{Discussion}
We have described a method for modelling the trip-based movements of animals undertaking central place foraging. This approach uses a hidden Markov model to directly estimate state movement and switching parameters from empirical telemetry observations. The model handles the natural sequence of behaviours within a trip, i.e.\ ``outbound'', ``search'', ``forage'', and ``inbound''.

\subsection{Comparisons to simpler models}
The four-state HMM we have constructed is relatively complex: it mixes biased and correlated random walks, and the transition probabilities depend on time-varying covariates (distance from colony, and time). It is therefore necessary to consider what we gain from using a complex behavioural model over simpler models. For example, two-state switching CRW models have commonly been used \citep[e.g][]{morales2004extracting, hindell2016circumpolar}, and can be easily fitted across a range of datasets, e.g.\ using the R packages moveHMM \citep{michelot2016movehmm} or bsam \citep{jonsen2016joint}. 

One key inference from the more complex model described in this manuscript is the length of outbound and inbound journeys (both in distance and in time). In this trip-based HMM, we can estimate this directly with the most likely state sequence, derived with the Viterbi algorithm. Arguably, one could apply heuristic rules to the state estimates from a two-state model, to obtain the same thing. For example, the outbound trip could be taken to start when the animal leaves the colony, and end when the proportion of observations categorized as ``resident'' behaviour reaches a threshold $p$ (where $p$ is relatively small, say $p=0.1$). This would have the effect of ignoring short runs of resident behaviour within the transit. Similar rules could be envisaged based on distance from the island, for example. These heuristic rules of thumb may be useful, but suffer from a degree of arbitrariness.

Hidden Markov models with more than two states have been used to model fishing vessel trips -- which can be considered, most basically, as another top predator. For example, \cite{vermard2010identifying} and \cite{walker2010pioneer} used 3-state HMMs to distinguish ``fishing'', ``steaming'', and ``still'' behaviours of fishing vessels. \cite{peel2011hidden} consider a 5-state HMM, with the addition of states for ``entry'' and ``exit'' movement between the latter two behaviours. In such models, simulation from fitted models could be a useful extension. However, to the best of our knowledge, they have not been used for that purpose. 

The disadvantage of our trip-based 4-state model is its aforementioned complexity. Without reasonable starting values in the maximum likelihood estimation, the parameter estimation routines can provide poor parameter estimates, by finding local maxima of the likelihood, or fail to converge altogether. These are well-known problems with numerical maximum likelihood, which require careful attention. One way to address this numerical problem is to run the estimation with many different sets of starting values (possibly chosen at random), and compare the resulting estimates. In the case study, we tried 50 sets of randomly chosen starting values in order to ensure that we identified the global maximum of the likelihood function.

\subsection{Progress toward prediction from estimated process models}
\label{S:prediction}
A key feature of the models we have demonstrated is that they are able to generate simulated tracks which capture certain aspects of the behaviour of seals. While it is clear that these are gross simplifications of the true movements of CPF predators, the simulations are nevertheless useful for predicting aggregate properties from the fitted model. For instance, we might predict the average spatial distribution of seals from the fitted model. We can also compute a distribution of arrival and departure times from a given area, which can be useful for assessing effectiveness of spatial management regimes or reserve usage and connectivity between populations \citep{stehfest2015markov,abecassis2013model,guan2013impacts,kanagaraj2013using}.

Currently, these models do not contain detailed environmental or biological predictors, which are known to be important in influencing southern elephant seal behaviour \citep{bestley2013integrative,pinaud2005scale}; for instance, the role of specific oceanographic variables \citep{biuw2007variations,labrousse2015winter}. It is in principle straightforward to include additional covariates in the model described in Section \ref{S:Model}, though doing so might increase numerical instability. Nonetheless, incorporating such variables will be important if these models are to truly realise their potential for understanding how CPF marine predators might respond to changing environmental regimes. For the case of southern elephant seals, we could for instance express the transition from outbound to search in terms of distance to the sea ice edge, instead of distance from the colony. We could also investigate the apparent state 2 behaviour observed further from the Antarctic continent, which may well be indicative of the animal moving slowly toward the colony and away from the continent as the ice advances northwards. Other variables, such as response to different water masses, frontal zones, etc., are likely to be more subtle, and may serve to influence transitions between search and forage states.

Including environmental covariates to inform the probabilities of switching may also lead to more realistic simulated tracks. In particular, it could help to simulate more realistic state dwell-times. An alternative is to use so-called hidden semi-Markov models \citep{langrock2012flexible}, where the geometric state dwell-time distribution can be replaced by more flexible distributions. In simulations, additional covariates would also help to inform the spatial distribution of foraging activities.

The approach presented here demonstrates progress toward melding telemetry and sensor data with spatially explicit prediction of animal distributions and behaviour. Hidden Markov and state space models have much to offer in this prospect \citep{patterson2008state}, but have thus far been rather limited in being used for the purposes of prediction. A long-term goal for animal movement research is the general prediction of realistic movements modelled (i.e.\ statistically estimated from a process model) from empirical data collected at the individual level, but applied to novel situations and scaled up to population-level responses. These might for example include projections of future environmental conditions \citep{perry2005climate,trathan2007environmental,hazen2013predicted}, or application to changed colony conditions.  

For the ultimate goal of building empirically and mechanistically based simulation models to be realised, we believe that it is necessary to directly estimate process models which capture the key aspects of animal biology sufficiently well. Recent studies have demonstrated that highly complex simulation models incorporating physiological details, habitat information, etc., can be built  \citep{schick2013estimating,new2014using}. However, typically such models are either data limited or unable to be directly estimated from empirical observations. As such, it is likely that there will be limitations to the degree of complexity which can be realised in estimated models. The consequence of this is that attempting to cleanly move from an estimated model to a simulation and prediction exercise will encounter difficulties as the estimation model fails to capture certain fundamental aspects of the real movements. A simple example of this is the behaviour of marine animals in regard to coastlines. Arbitrary, but probably reasonable measures, such as using a rejection step to restrict animals to remain in the ocean are necessary to mimic the straightforward reality that marine animals do not typically wander over land masses. Despite the apparent triviality of this point, it is informative to consider, as it highlights elements needed at the simulation and prediction phase, but which may not fit within an estimation model.

\subsection{Concluding remarks}

Building on the general framework of Markov-switching random walks and hidden Markov models, our method accommodates naturally trip-based movement of central place foragers. It offers a fast way to categorize movement tracks into behavioural modes, and to describe the underlying mechanics of behavioural switching in terms of time-varying covariates. We believe that models like those presented here begin to address the interesting three way trade-off between (1) complexity and realism, (2) the desirable aspects of direct estimation using rigorous statistical inference, and (3) computational efficiency.  The first aspect allows simulations to capture many features of the real data and makes the models potentially useful for prediction at the individual level. The second brings the power and objectivity of statistical methods as a way to understand the spatial dynamics of animals. The final point allows for ease of use, and means that more realistic models can be applied to large data sets. 

\renewcommand{\baselinestretch}{1}

\section*{Acknowledgments}
{\small TM and TP received support from  IMBER-CLIOTOP and Macquarie University Safety Net Grant 9201401743. SB was supported under an Australia Research Council Super Science Fellowship. IDJ was supported by a Macquarie Vice-Chancellor’s Innovation Fellowship. TAP was supported by a CSIRO Julius career award and the Villum foundation. The southern elephant seal data was sourced from the Integrated Marine Observing System (IMOS). The tagging program received logistics support from the Australian Antarctic Division and the French Polar Institute (Institut Paul-\'Emile Victor, IPEV). All tagging procedures were approved and carried out under the guidelines of the University of Tasmania Animal Ethics Committee and the Australian Antarctic Animal Ethics Committee. Initial ideas from Mark Bravington, Uffe Thygesen and Martin Pedersen were very helpful in formulating earlier versions of these models. We thank an anonymous reviewer for helpful feedback on an earlier version of the manuscript.}

\bibliographystyle{apalike}
\bibliography{refs.bib}

\section*{Appendix}

\subsection*{A1. State-decoded tracks}
\includegraphics[width=0.49\textwidth]{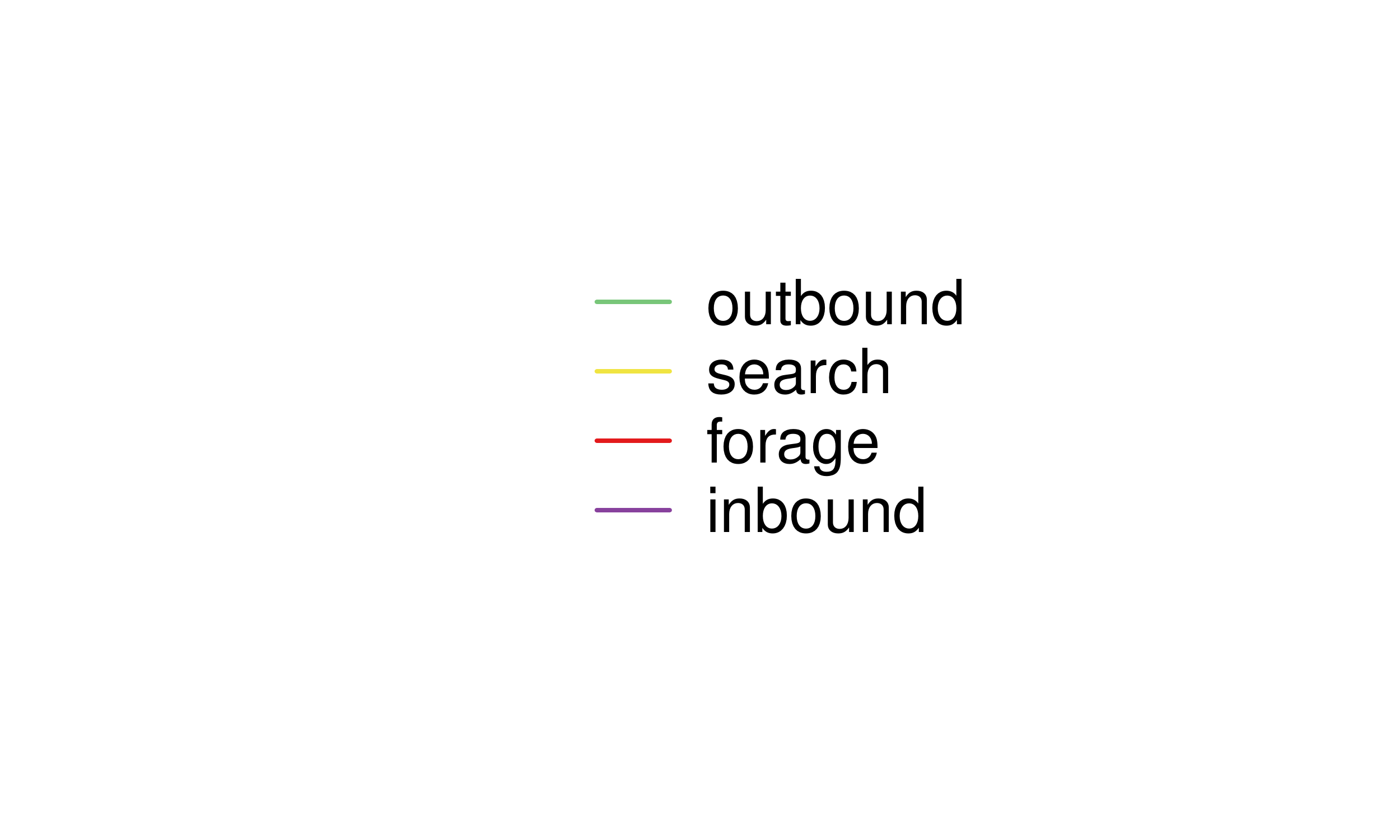}
\includegraphics[width=0.49\textwidth]{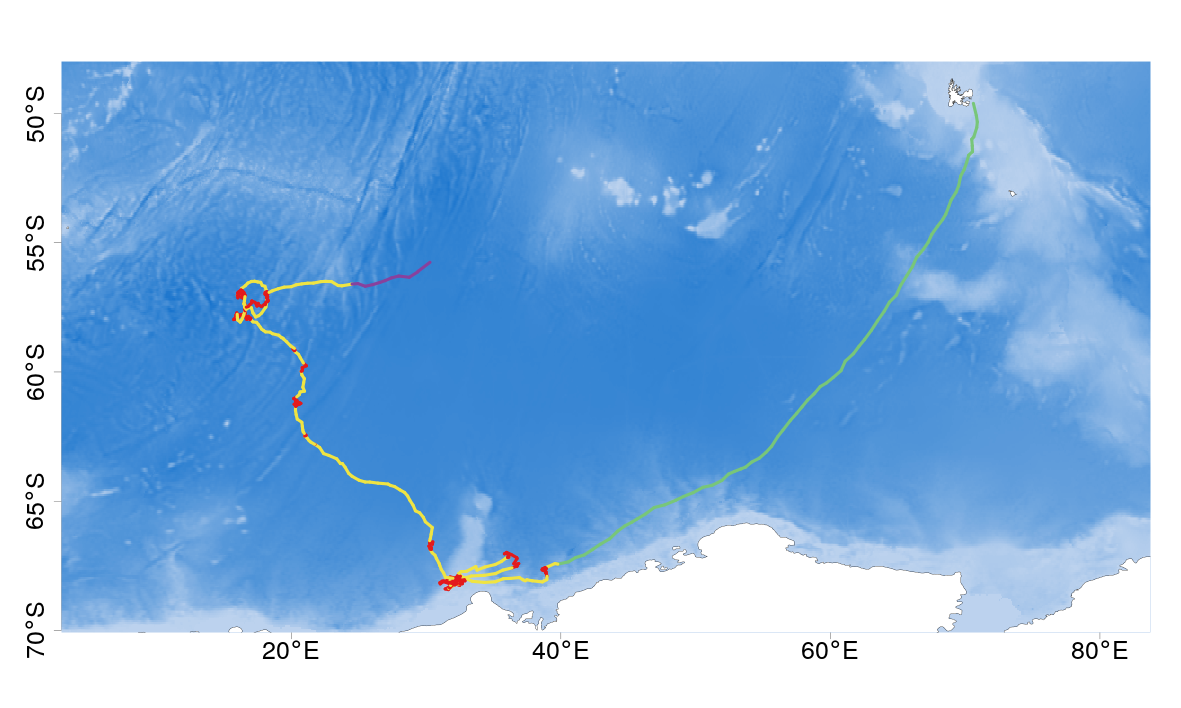}\\
\includegraphics[width=0.49\textwidth]{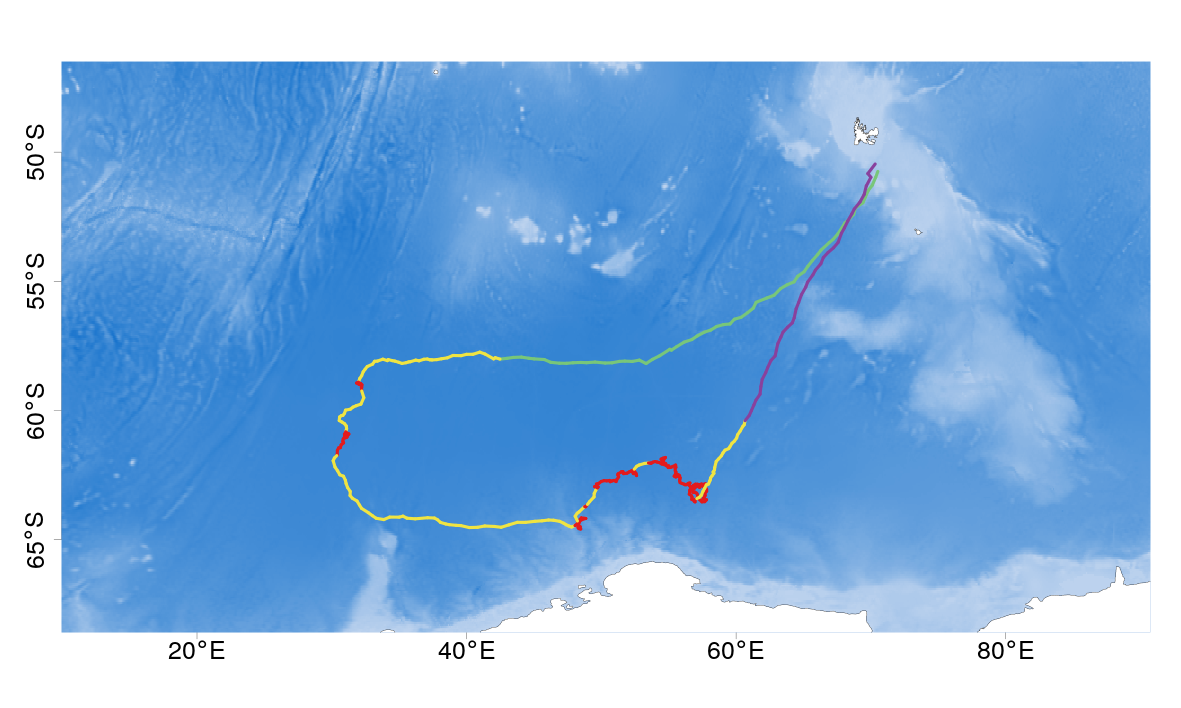}
\includegraphics[width=0.49\textwidth]{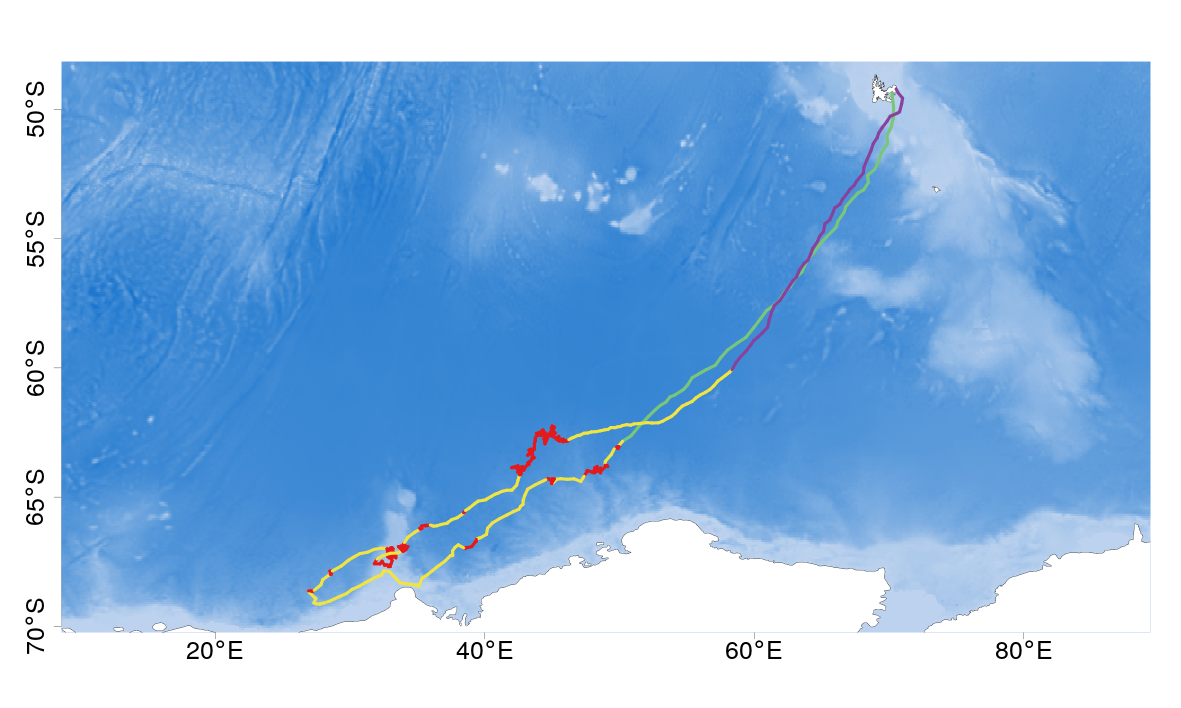}\\
\includegraphics[width=0.49\textwidth]{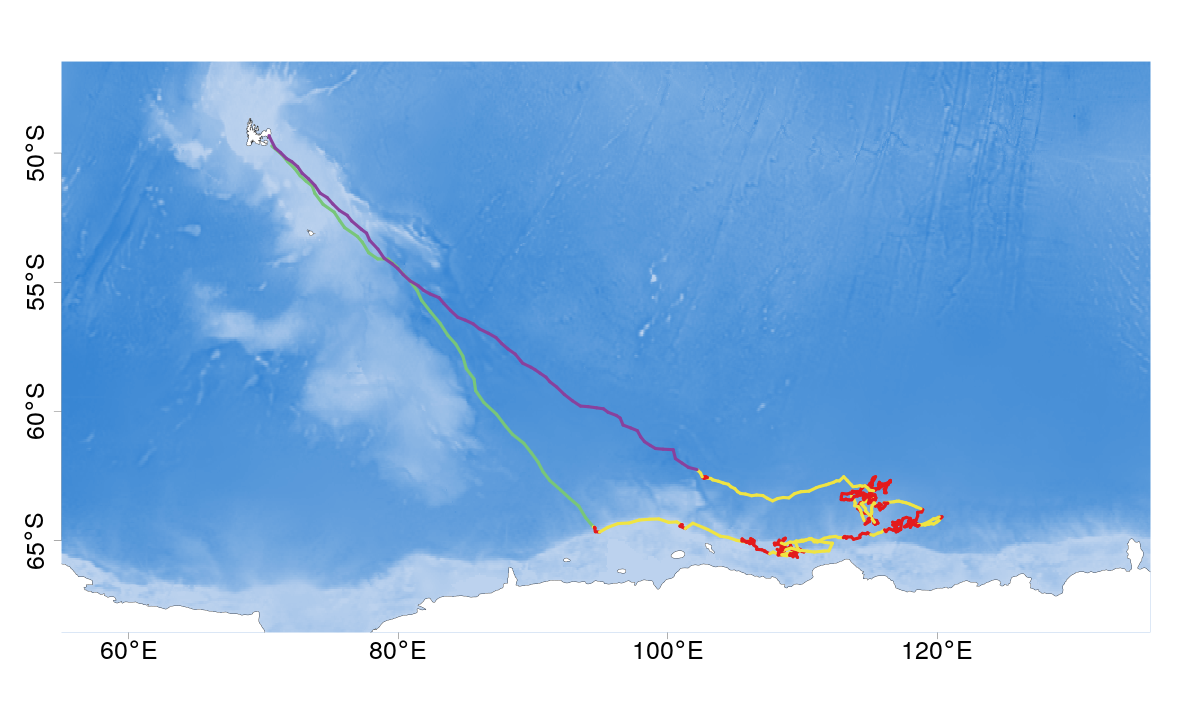}
\includegraphics[width=0.49\textwidth]{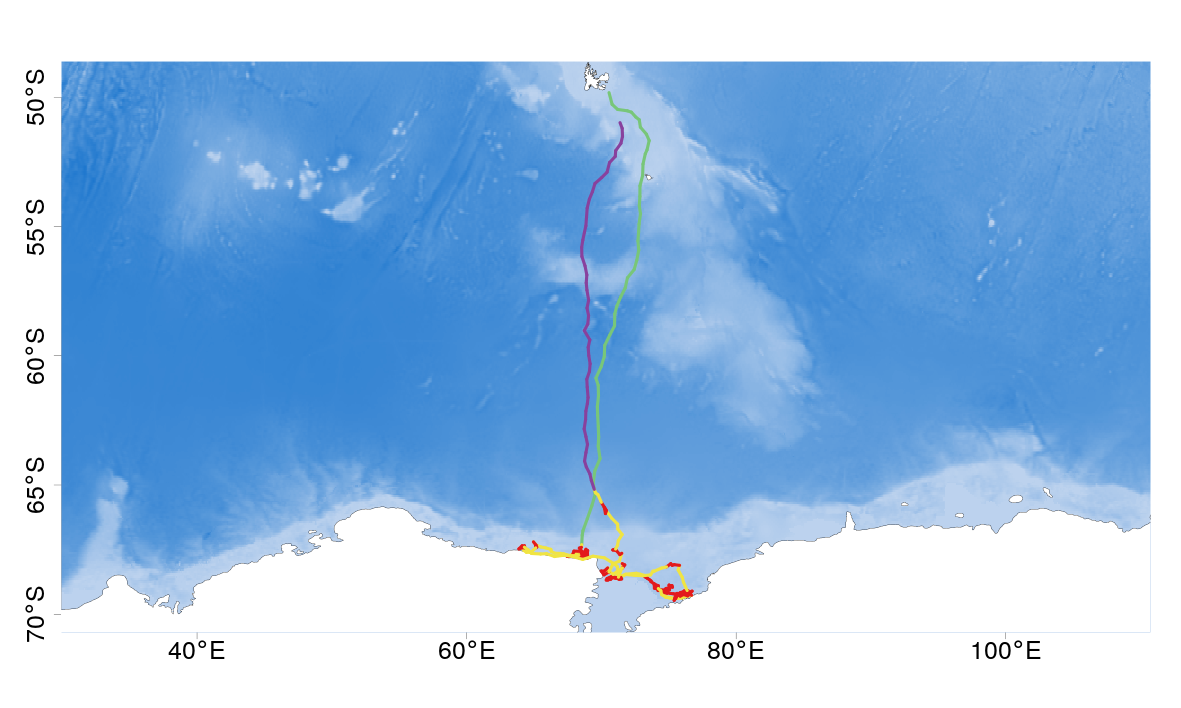}\\
\includegraphics[width=0.49\textwidth]{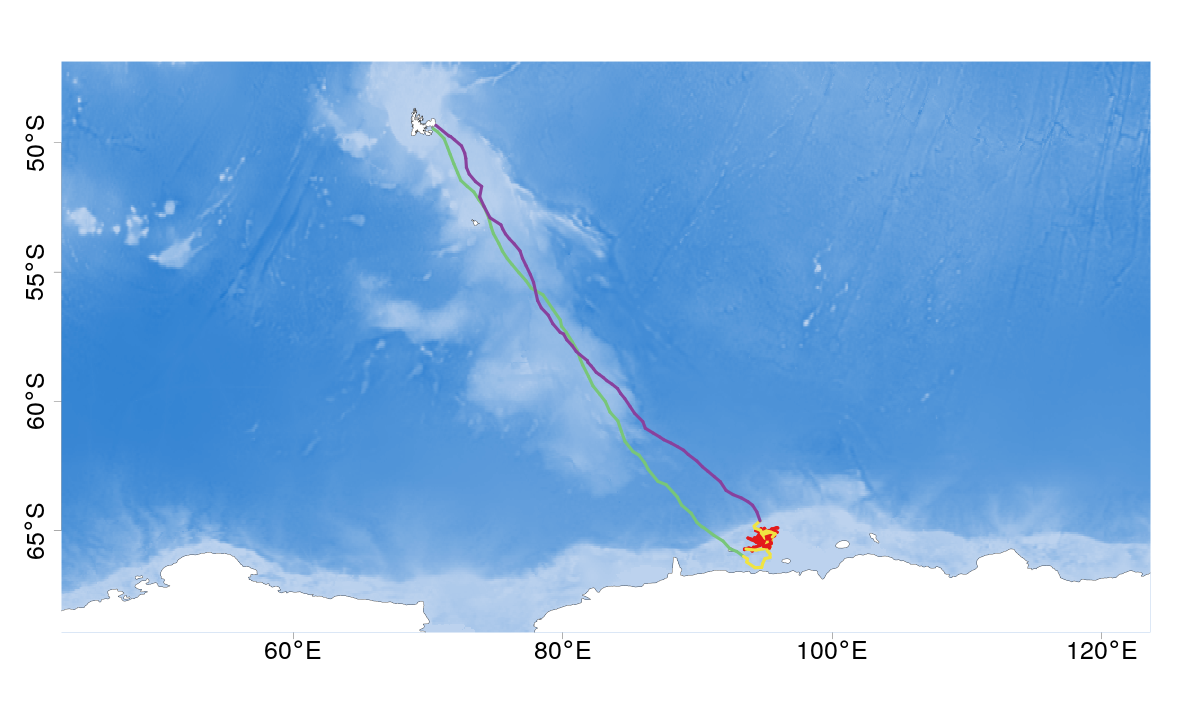}
\includegraphics[width=0.49\textwidth]{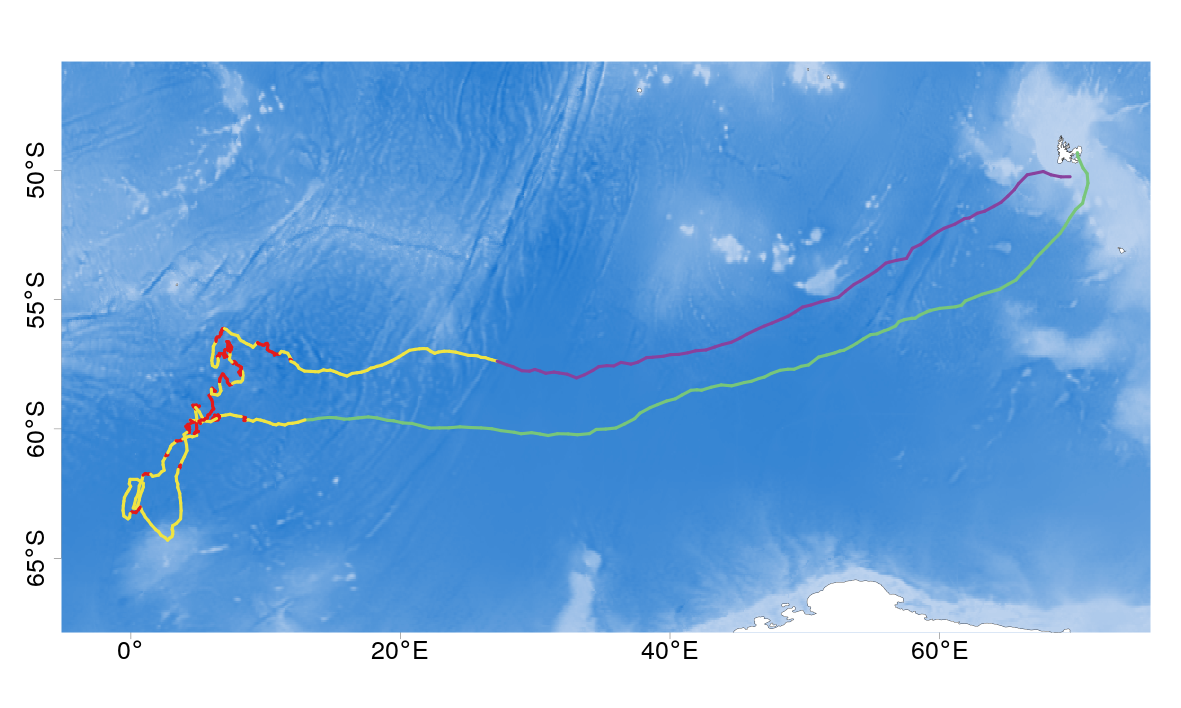}\\
\includegraphics[width=0.49\textwidth]{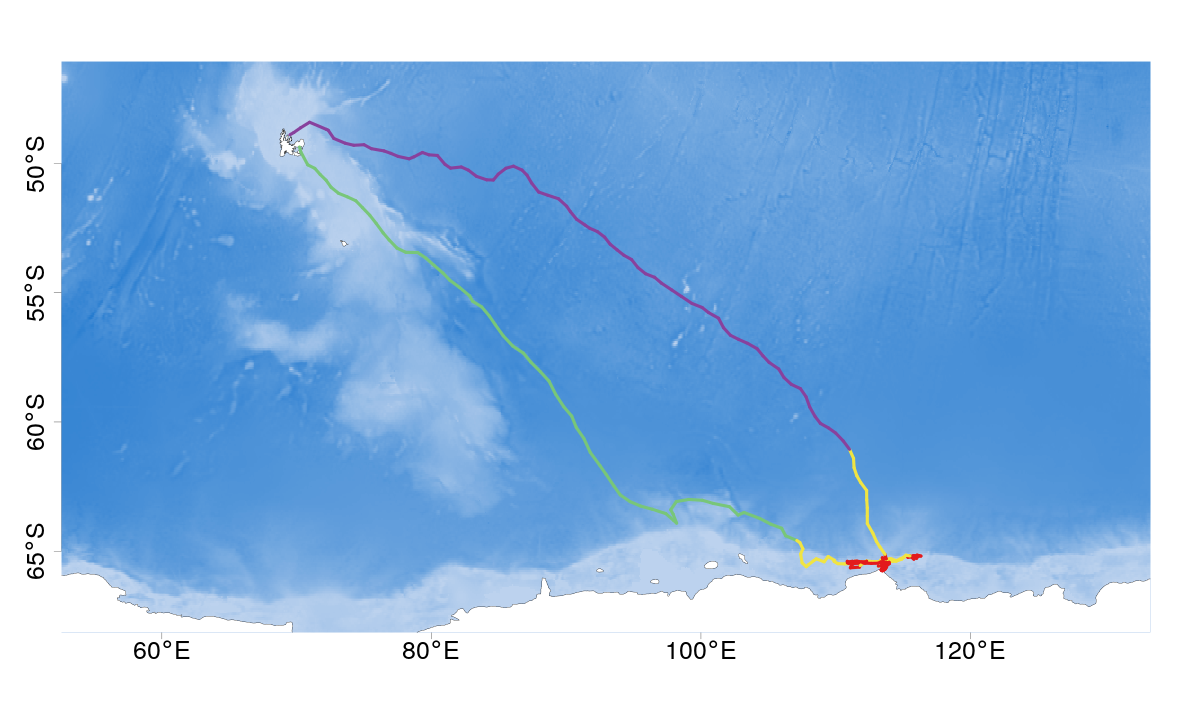}
\includegraphics[width=0.49\textwidth]{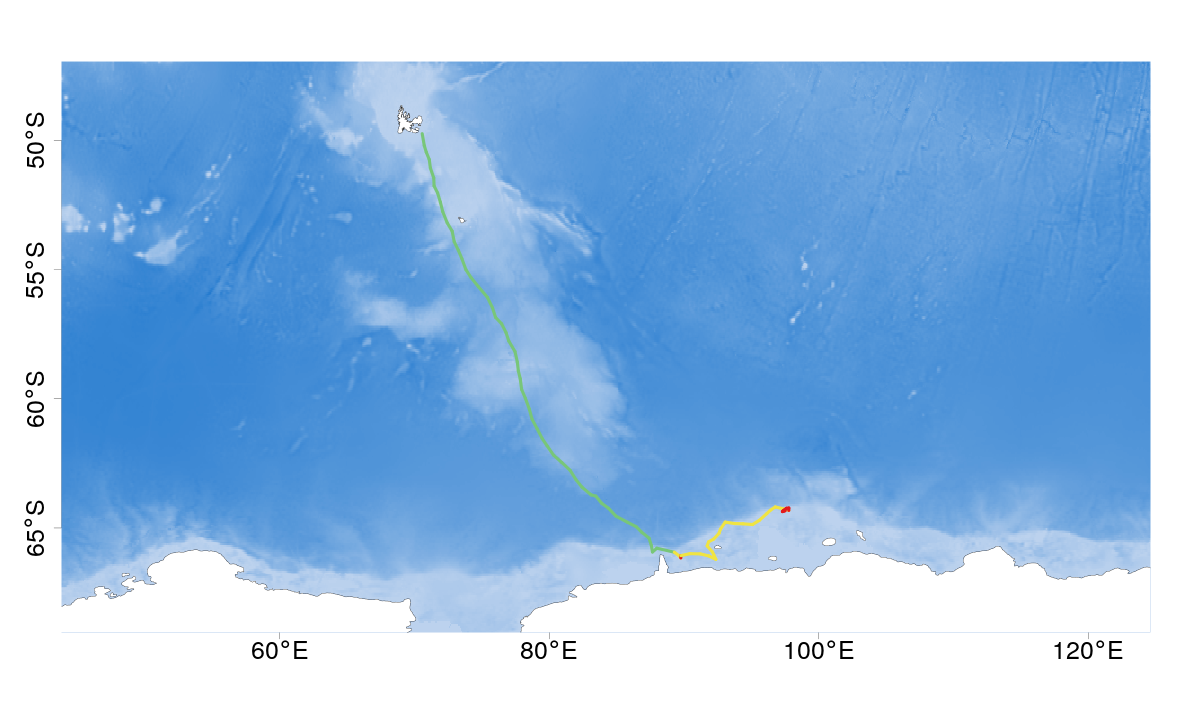}\\
\includegraphics[width=0.49\textwidth]{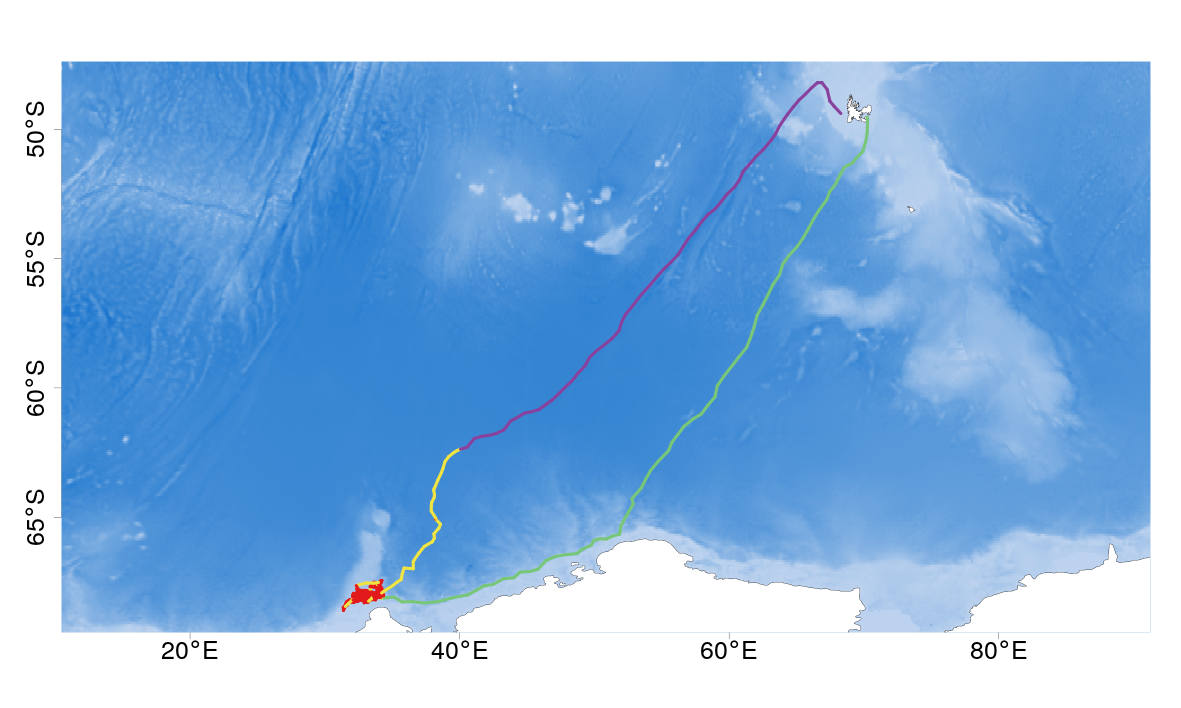}
\includegraphics[width=0.49\textwidth]{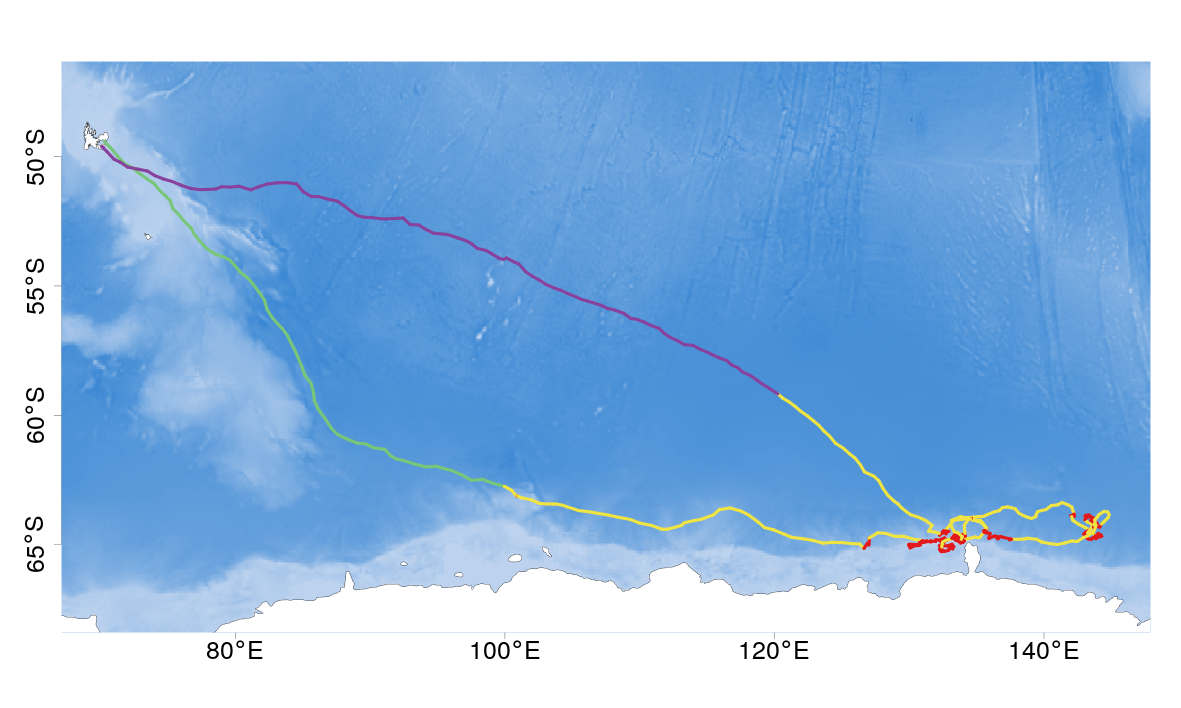}\\
\includegraphics[width=0.49\textwidth]{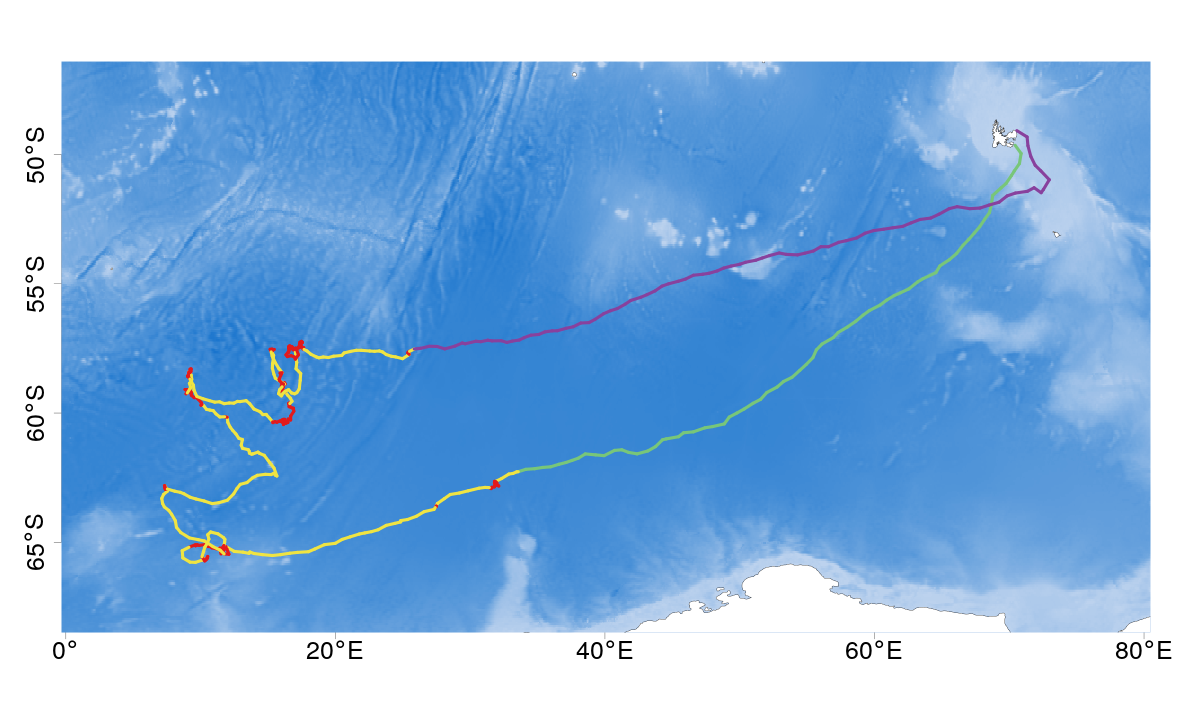}
\includegraphics[width=0.49\textwidth]{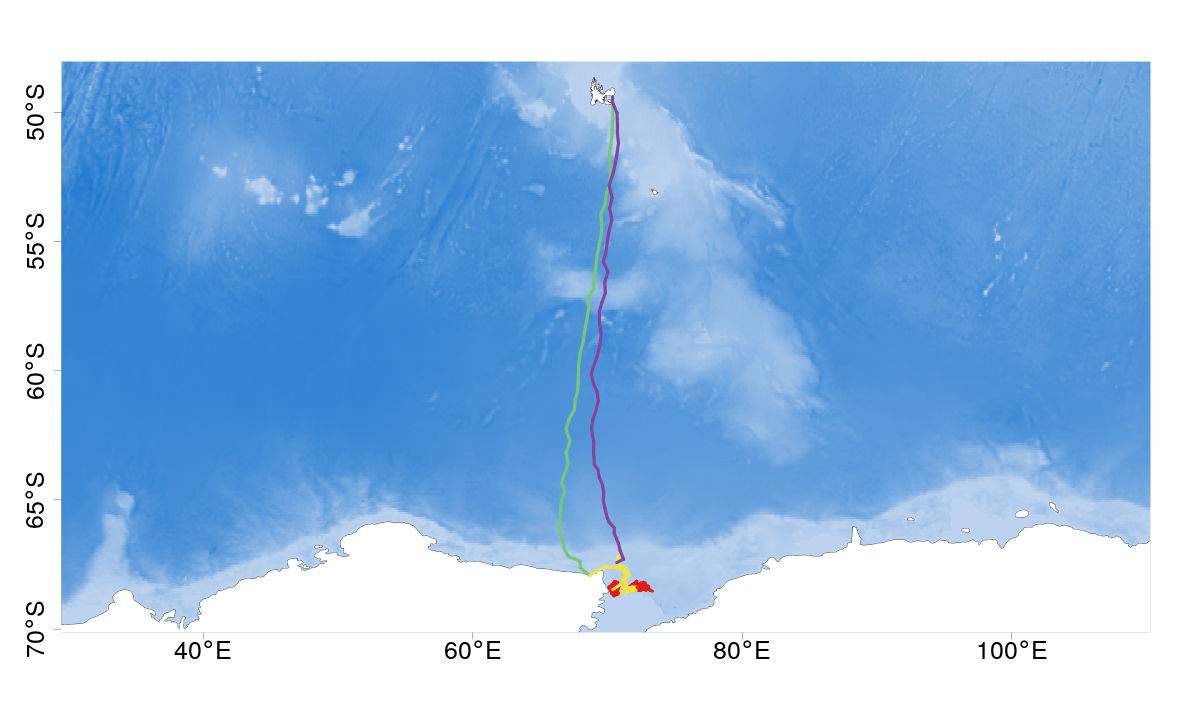}\\
\includegraphics[width=0.49\textwidth]{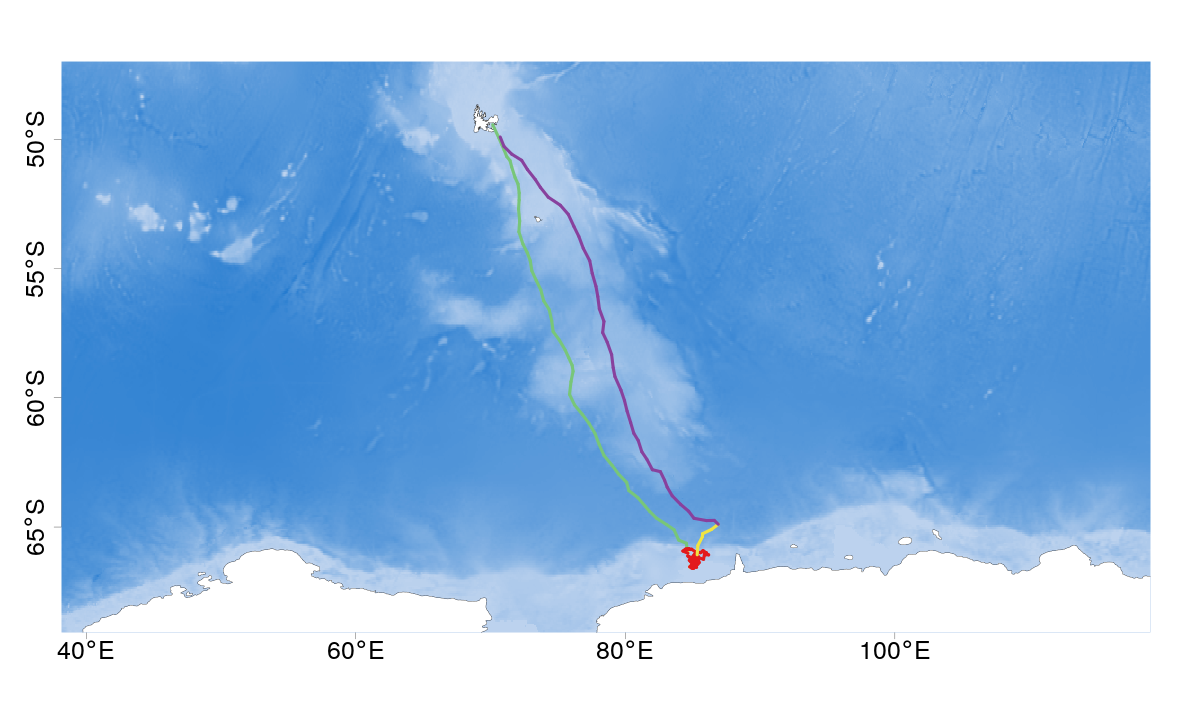}
\includegraphics[width=0.49\textwidth]{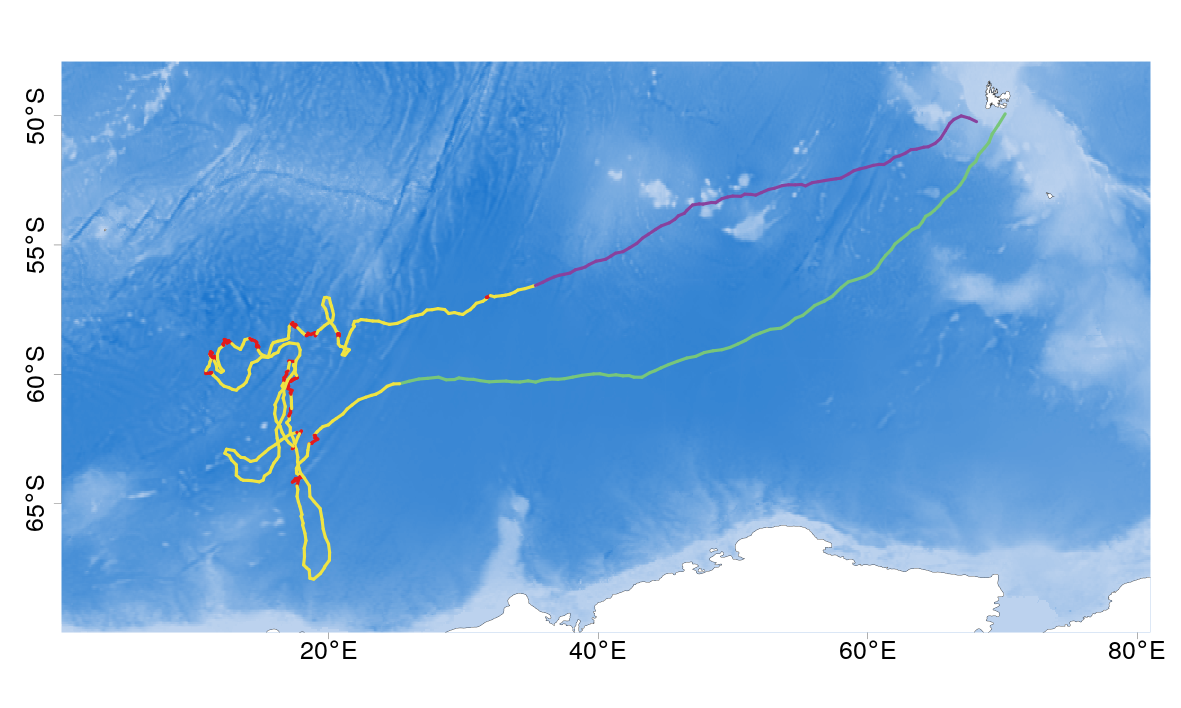}

\newpage
\subsection*{A2. Parameter estimates and CIs}
\begin{table}[htb]
  \centering
  \begin{tabular}{rccc}
    & & Estimate & 95\% CI \\
    \midrule
    $\mu_1$ & & 40.23 & [39.52,\ 40.94] \\[1mm]
    $\mu_2$ & & 20.95 & [20.39,\ 21.53] \\[1mm]
    $\mu_3$ & & 8.10 & [7.89,\ 8.31] \\[1mm]
    $\mu_4$ & & 38.04 & [37.25,\ 38.85] \\[1mm]
    $\sigma_1$ & & 10.96 & [10.43,\ 11.52] \\[1mm]
    $\sigma_2$ & & 8.80 & [8.44,\ 9.18] \\[1mm]
    $\sigma_3$ & & 5.30 & [5.12,\ 5.49] \\[1mm]
    $\sigma_4$ & & 11.09 & [10.51,\ 11.69] \\[1mm]
    $\lambda_2$ & & -0.01 & [-0.03,\ 0.02] \\[1mm]
    $\lambda_3$ & & 3.14 & --- \\[1mm]
    $\kappa_1$ & & 6.87 & [6.29,\ 7.51] \\[1mm]
    $\kappa_2$ & & 3.70 & [3.36,\ 4.08] \\[1mm]
    $\kappa_3$ & & 0.08 & [0.04,\ 0.16] \\[1mm]
    $\kappa_4$ & & 8.28 & [7.51,\ 9.12] \\[1mm]
    $\beta_0^{(12)}$ & & -7.26 & [-8.97,\ -5.54] \\[1mm]
    $\beta_1^{(12)}$ & & 1.67 & [0.98,\ 2.36] \\[1mm]
    $\beta_0^{(23)}$ & & -2.21 & [-2.40,\ -2.03] \\[1mm]
    $\beta_0^{(24)}$ & & -9.79 & [-12.79,\ -6.80] \\[1mm]
    $\beta_1^{(24)}$ & & 1.00 & [0.51,\ 1.48] \\[1mm]
    $\beta_0^{(32)}$ & & -3.00 & [-3.18,\ -2.82] 
  \end{tabular}
  \caption{Estimates of the model parameters and their 95\% confidence intervals. In state $i$, $\mu_i$ and $\sigma_i$ are the mean and standard deviation of the gamma distribution for the step lengths, respectively; $\lambda_i$ and $\kappa_i$ are the mean and the concentration of the von Mises distribution for the turning angles, respectively. The step lengths are measured in kilometres, and the angles in radians. The $\beta_{k}^{(ij)}$ are defined as in the model formulation, and $\beta_0^{32} = \text{logit}(\gamma_{32})$. The confidence intervals were derived from the Hessian of the log-likelihood function. The confidence interval for $\lambda_3$ is not given because the estimate is on the boundary of the parameter space, $(-\pi,\pi]$.}
\end{table}

\newpage
\subsection*{A3. Dwell time distributions}
\begin{figure}[htb]
  \centering
  \includegraphics[width=\textwidth]{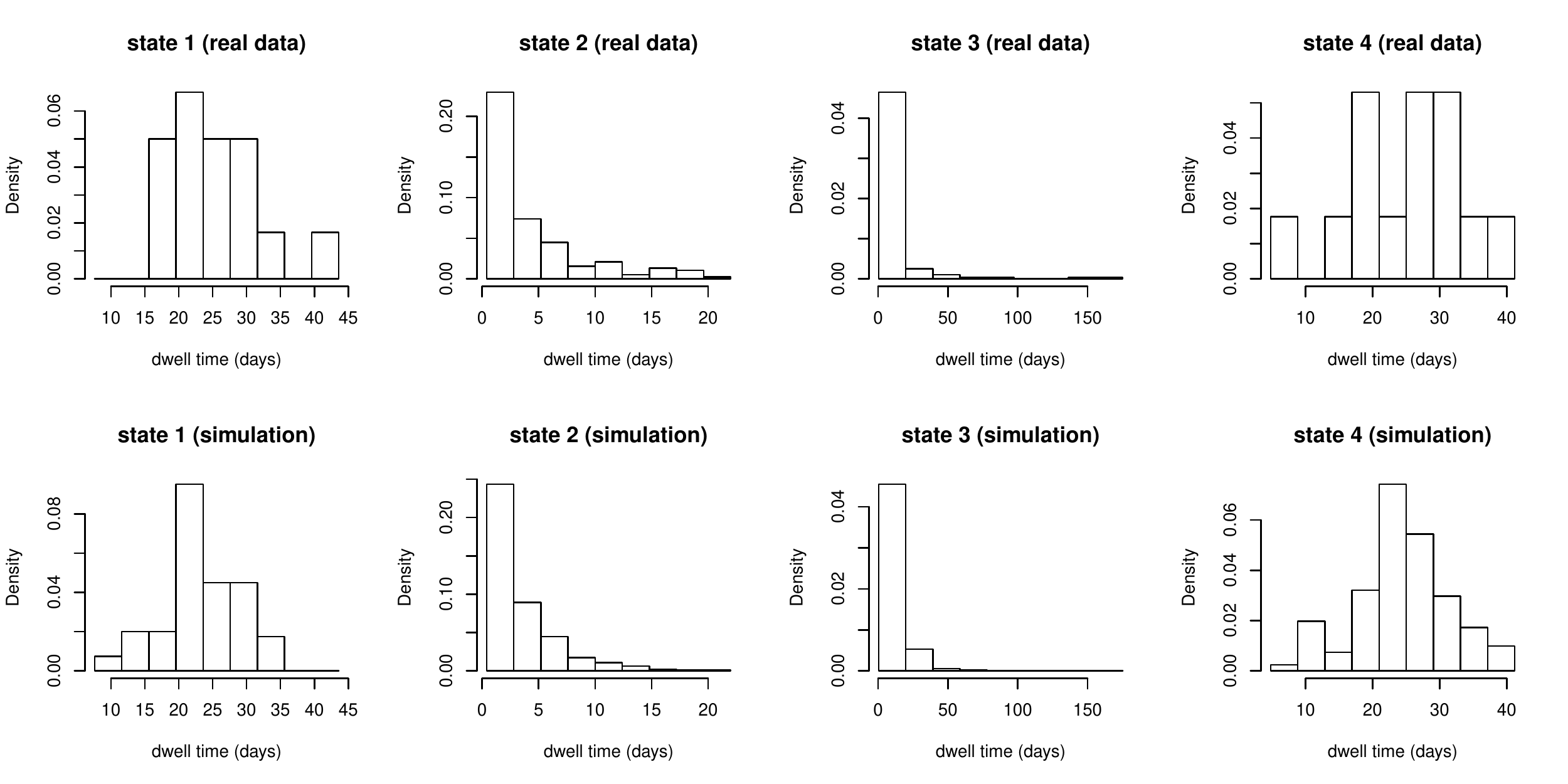}
  \caption{Histograms of the dwell times in each state, for the 15 real tracks (top row) and the 100 simulated tracks (bottom row).}
\end{figure}

\newpage
\subsection*{A4. Pseudo-residuals}
\begin{figure}[htb]
  \centering
  \includegraphics[width=0.7\textwidth]{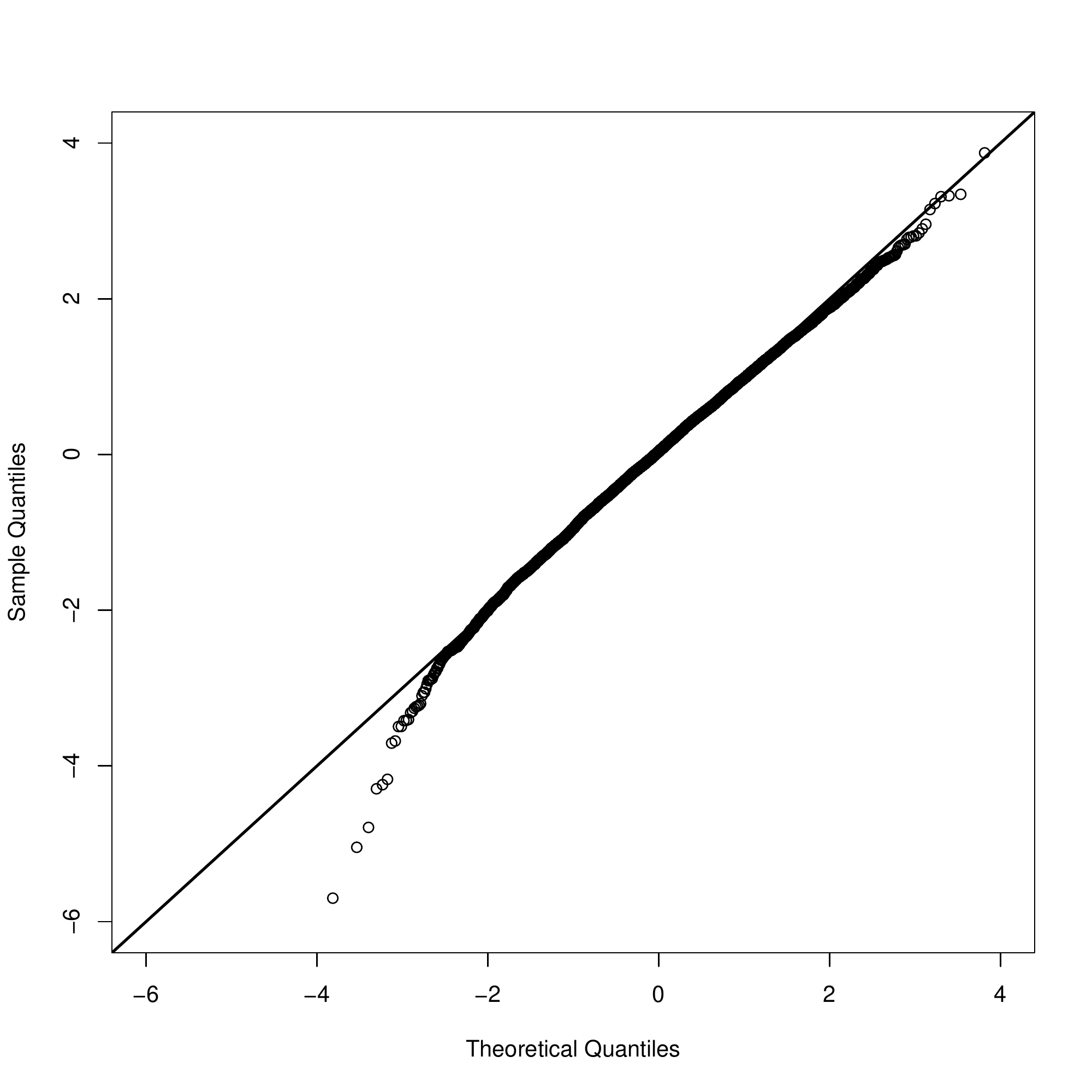}
  \caption{Standard normal qq-plot of the pseudo-residuals for the step lengths.}
\end{figure}

\end{document}